\documentclass[sigplan, nonacm]{acmart}
\newcommand{\asplossubmissionnumber}{220}

\usepackage[normalem]{ulem}
\usepackage[english]{babel}
\usepackage{amsthm}

\usepackage{color}
\usepackage[utf8]{inputenc}
\usepackage{fancyvrb}
\usepackage{xcolor}

\usepackage{caption}
\usepackage{subcaption}

\usepackage{multirow}

\newcommand{\fixit}[1] {
  {\color{red}{Eddy: #1}}}
\newcommand{\xiangyu}[1] {
  {\color{blue}{Xiangyu:#1}}}

\theoremstyle{definition}

  \newcommand{\etal}{\textit{et al}.}

\begin{document}

\title{Quantum Fourier Transformation Circuits Compilation}
\author{Yuwei Jin$^{1*}$,  Xiangyu Gao$^{2*}$, Minghao Guo$^{1}$, Henry Chen$^{1}$, Fei Hua$^{1}$, Chi Zhang$^{3}$, Eddy Z. Zhang$^{1}$}
\author{$^{1}$ Rutgers University $^{2}$ New York University $^{3}$ University of Pittsburgh}


\begin{abstract}

In this research paper, our primary focus revolves around the domain-specific hardware mapping strategy tailored for Quantum Fourier Transformation (QFT) circuits. While previous approaches have heavily relied on SAT solvers or heuristic methods to generate hardware-compatible QFT circuits by inserting SWAP gates to realign logical qubits with physical qubits at various stages, they encountered significant challenges. These challenges include extended compilation times due to the expansive search space for SAT solvers and suboptimal outcomes in terms of the number of cycles required to execute all gate operations efficiently.

In our study, we adopt a novel approach that combines technical intuition, often referred to as "educated guesses," and sophisticated program synthesis tools. Our objective is to uncover QFT mapping solutions that leverage concepts such as affine loops and modular functions.

The groundbreaking outcome of our research is the introduction of the first set of linear-depth transformed QFT circuits designed for Google Sycamore, IBM heavy-hex, and the conventional 2-dimensional (2D) grid configurations, accommodating an arbitrary number of qubits denoted as 'N'. Additionally, we have conducted comprehensive analyses to verify the correctness of these solutions and to develop strategies for handling potential faults within them.\footnote{ * These authors contributed equally to this work.}

\end{abstract}
\maketitle
\pagestyle{plain}

\thispagestyle{empty}

\section{Introduction}

Quantum computing is gaining traction in recent years because of its tremendous computation power in accelerating certain types of applications including cryptography \cite{shor:focs94}, database search \cite{grover+:stoc96}, and chemistry simulation \cite{peruzzo+:ncomms14}. 

Quantum Fourier Transformation (QFT) is an important algorithm in quantum computing. It is at the heart of Shor's algorithm. Shor's algorithm is designed for factoring large numbers into prime factors. Shor's algorithm has a significant impact on cryptography as widely used RSA algorithms rely on the difficulty of factoring large numbers. Shor's algorithm has exponential speedup \cite{shor:focs94} over the state-of-the-art classical factoring algorithm thanks to QFT. 

In this paper, we focus on how to realize QFT on modern quantum architectures with connectivity constraints. Today's quantum hardware typically has limited connectivity. Superconducting qubits use Joseph Junctions (JJ) transmons, and the connectivity between JJ units are semi-conductor couplers. The number of couplers is typically sparse and linear with respect to the number of qubits. 
In contrast, the theoretical design of QFT assumes the hardware has all-to-all connectivity, which is not true in today's real devices. Hence, we need to map QFT to real hardware with connectivity constraints efficiently.

\vspace{-7pt}

\begin{figure}[htb]
    \centering
    \includegraphics[width=0.45\textwidth]{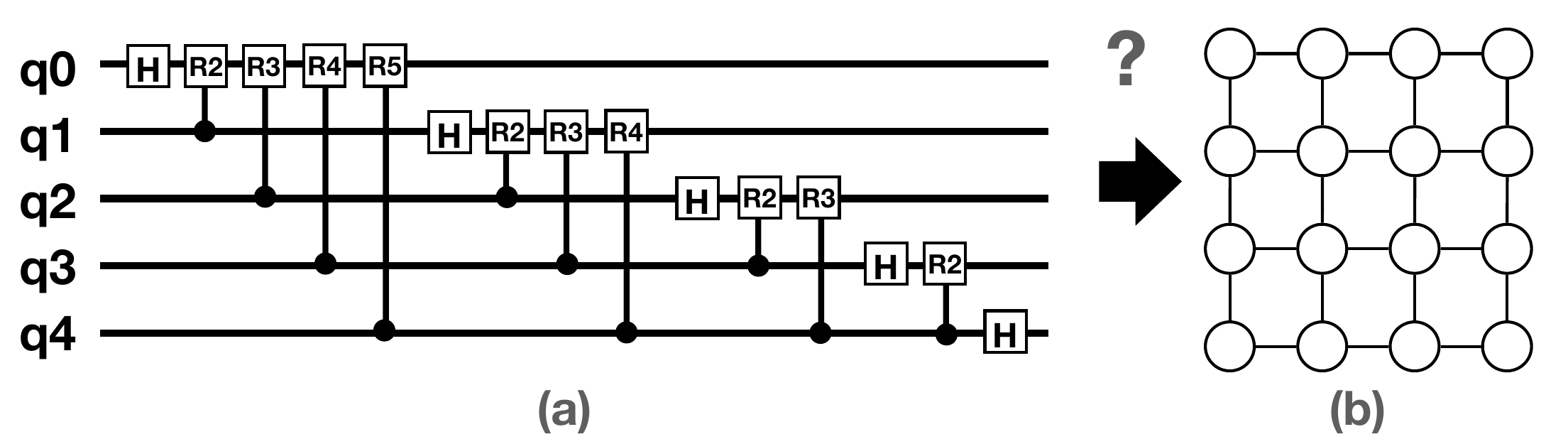}
    \vspace{-8pt}
    \caption{(a) QFT logical circuit with all-to-all interaction pattern. Each wire represents a qubit.  (b) A non-trivial two dimensional architecture that QFT needs to be mapped to.}
    \label{fig:qftlogical}
\end{figure}

\paragraph{Overview of the QFT Mapping Problem}
The QFT algorithm has an all-to-all qubit communication pattern. The structure of a QFT circuit is shown in Fig. \ref{fig:qftlogical} (a). A qubit needs to interact with every other qubit using a two-qubit gate (CPHASE - controlled phase rotation). There is a dependence order between these two-qubit gates. QFT represents one of the most challenging types of applications for hardware mapping: from all-to-all gate interaction to an architecture with a linear number of hardware links. For instance, in a 2 dimensional (2D) grid architecture, the number of hardware links is less than $4N$, where $N$ is the number of qubits. 

\vspace{-5pt}

\vspace{-2pt}

QFT is a computation kernel that is used beyond Shor's algorithm. It serves as the basis for a number of efficient quantum algorithms. QFT is used in quantum phase estimation, quantum state amplitude amplification, data compression, quantum linear algebra, chemistry and simulation, as well as quantum cryptography that involves quantum key distribution (QKD), and etc. Moreover, QFT's circuit structure bears a resemblance to other non-trivial applications' structures in quantum computing. For instance, the structure of QFT shown in Fig. \ref{fig:qftlogical} (a) is similar to that of the stabilizer circuits (known as unitary stabilizer circuits or Clifford group circuits) \cite{maslov:physreva16}\cite{aaronson+:physreva04}  \cite{bennett+:physreva96, gottesman+:physreva96}. Hence, if we can map QFT efficiently to hardware, we can map a number of important quantum applications to hardware efficiently.

\begin{figure}[htb]
    \centering
\includegraphics[width=0.3\textwidth]{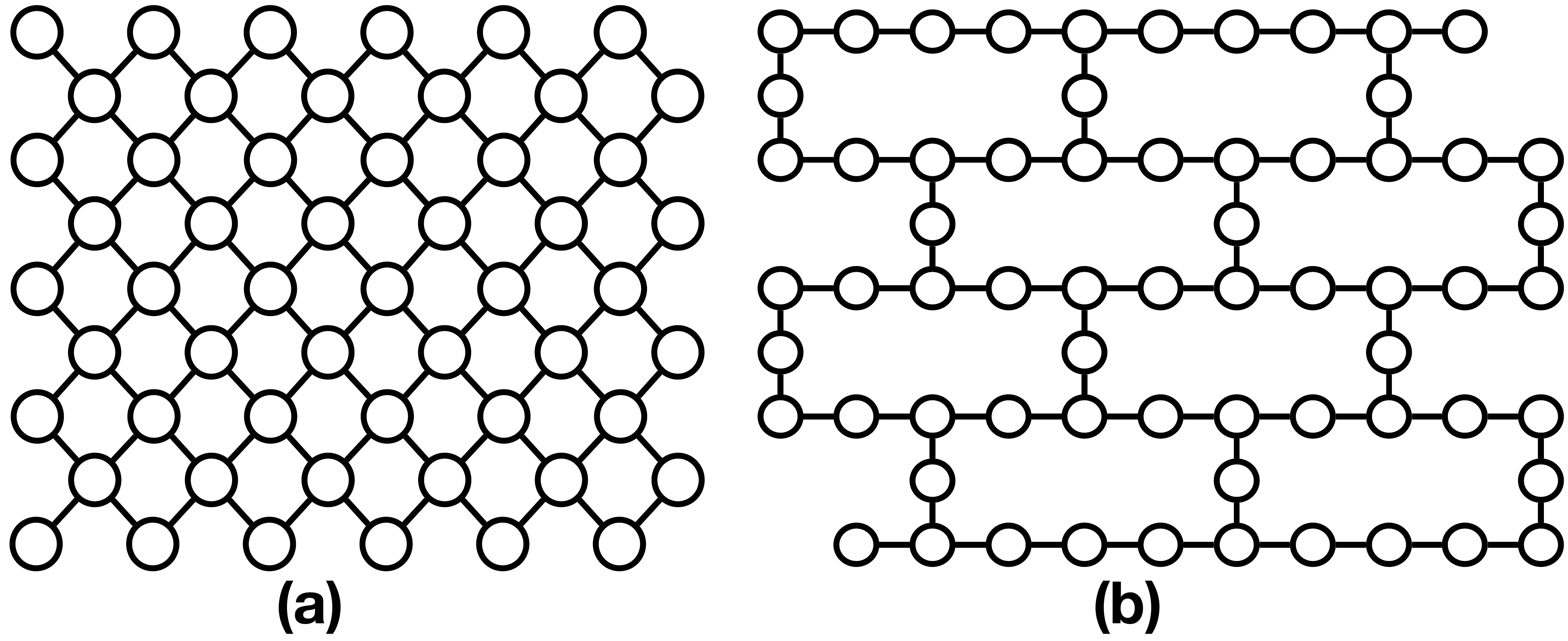}
\vspace{-7pt}
    \caption{(a) Google Sycamore (b) Heavy-hex}
    \label{fig:modernarch}
    \vspace{-12.5pt}
\end{figure}

\textbf{\emph{Prior Work}}
Most prior studies perform qubit mapping for a general class of applications \cite{siraichi+:cgo18, li+:asplos19, zhang+:asplos21, molavi+:micro22, tan+:iccad20, tan+:iccad22}. Domain-specific qubit mapping approaches exist but only for a small set of applications, for instance, for quantum approximate optimization algorithms (QAOA) \cite{alam+:dac20, alam+:micro20, lao+:isca22} and variational quantum eigensolvers (VQE) \cite{li+:asplos22, li+:isca21}. 
For QFT, Maslov \etal  ~\cite{maslov+:physreva07} for the first time demonstrates a linear time solution on the linear nearest neighbor (LNN) architecture. LNN represents a path of connected qubits. However, it is difficult to find a Hamiltonian path that connects all nodes in modern quantum architectures. Examples are shown in Fig. \ref{fig:modernarch} (a) and (b), respectively for Google's and IBM's superconducting architecture.   Zhang \etal~ \cite{zhang+:asplos21} improved upon Maslov's \etal~ \cite{maslov+:physreva07} by discovering a linear-depth solution for a 2D grid with only two rows. However such architecture with only two rows does not exist in scalable modern architectures. 

\vspace{-5pt}

\vspace{-5pt}

\paragraph{Our Contributions}
Our contributions in this paper are multi-fold. We, for the first time, discovered linear-depth QFT  mapping solutions for Google Sycamore, IBM heavy-hex architecture, and a general 2D grid architecture. In addition to discovering linear-depth solutions for mapping QFT to these modern multi-dimensional architectures, we also demonstrate our experience of using program synthesis \cite{sketch} methods to integrate human intelligence in order to find such structured solutions (faster). Our experience and methodology for using program synthesis in qubit mapping  is not only useful for QFT, but may potentially useful for other quantum programs and architecture with a regular structure. Our contributions of this paper can be summarized as follows: 

\begin{itemize}
    \item Linear-depth solutions for QFT on different architectures: 2D grid, Google Sycamore, and IBM Heavy-hex.
\item Program synthesis formalism for finding QFT mapping solutions on different regular architectures
    \item Technical intuition such as a divide-and-conquer approach and an in-depth understanding of the LNN solution for tackling more complex architectures.  

    \item Our QFT mapping solutions have up to $52\%$ fewer SWAP gate count and $95\%$ fewer depth than state-of-the-art approaches for up to 1600 qubits.
    
    \item Fault (or error-prone qubits) consideration for a regular architecture when adapting the analytical QFT hardware mapping solutions.  
\end{itemize}

\section{Background and Motivation}
\label{sec:motivation}
We will describe the definition of the hardware mapping problem, and then introduce our observations that drive us to further study the mapping of QFT algorithm on modern hardware via program synthesis approaches.

\begin{figure}[htb]
    \centering
\includegraphics[width=0.33\textwidth]{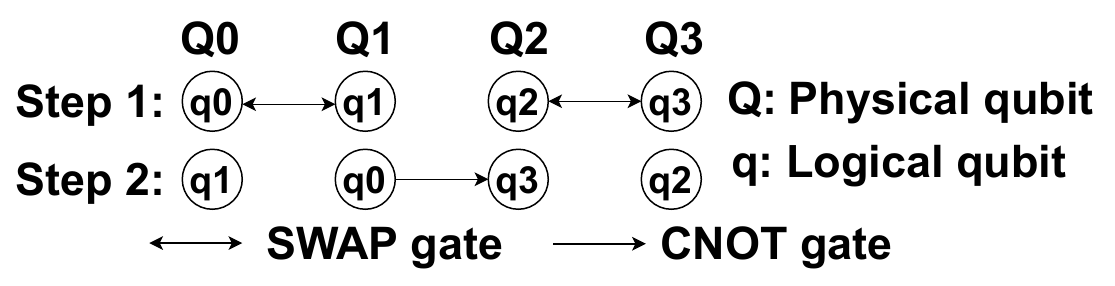}
    \vspace{-5pt}
    \caption{SWAP insertion to implement CNOT(q0, q3).}
    \label{fig:connectivity}
    \vspace{-0.1in}
\end{figure}

{\subsection{Hardware Mapping}}
In superconducting hardware, a two-qubit gate cannot be executed until its two qubits are located in 2 connected physical qubits. In reality, due to the sparse connectivity inside the current quantum machine, SWAP gates are required to change the hardware mapping on the fly and make sure each two-qubit gate has their logical qubits physically connected right before the gate itself. For instance, in order to execute a two-qubit gate between logical qubit q0 and q3 in Fig.~\ref{fig:connectivity}, we can insert 2 SWAP gates to move q0 and q3 to physical qubits Q1 and Q2 respectively, making it possible to run a 2-qubit gate on them. 

\vspace{-5pt}

\vspace{-5pt}

\subsection{Program Synthesis}
A program synthesis process requires a specification and an implementation. The specification describes what needs to be achieved, for instance, we require all CPHASE and H gates in QFT to be executed with respect to its dependencies in the transformed circuit, and also SWAPs are inserted for hardware-compliance. The implementation specifies the shape of the code that can potentially achieve the goal in the specification. But such a code shape is roughly specified, and certain parameters remain to be solved. For instance, the loop bounds as an affine function, or a finite number of statement choices at a particular location of the code shape. An example is shown in Fig. \ref{fig:codesyn}.

The synthesizer will return two types of results. If there is a correct implementation that satisfies the specification, it outputs the concrete parameters and code choices. Otherwise, it will return false. 

\subsection{Hardware Mapping and Program Synthesis}

Prior work by Maslov \cite{maslov:physreva16} and Zhang \etal~ \cite{zhang+:asplos21} discovered the same solution for mapping QFT to the linear nearest neighbor (LNN) architecture. 

\begin{figure}[htb]
    \centering
\includegraphics[width=0.48\textwidth]{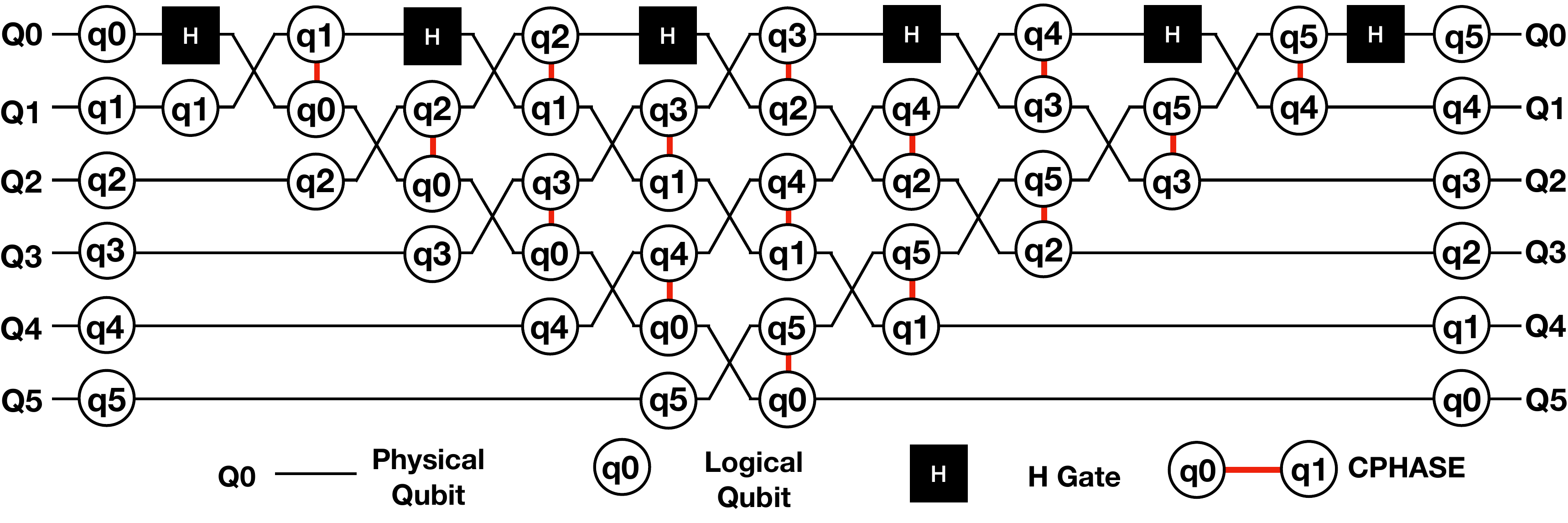}

    \vspace{-5pt}
    
    \caption{Hardware mapped QFT in LNN. $q_i$ is a logical qubit. $Q_i$ is a physical qubit. The single-qubit gate runs in parallel with two-qubit gates. Each qubit moves to the top first and then moves down, except q0, which directly moves down. When a qubit is at the top, it stops for one time step. }   
    
    \label{fig:LNN5}
    \vspace{-5pt}
\end{figure}

We show the LNN solution via a small example in Fig. \ref{fig:LNN5}.  The circuit clearly exhibits a pattern. Assuming N is the number of qubits, $q_i$ represents a logical qubit, and $Q_i$ represents a physical qubit.  The logical qubit $q_i$ is mapped to $Q_i$ in the initial mapping.  

We denote a CPHASE gate between physical qubits $Q_i$ and $Q_{i+1}$ as $G(Q_i, Q_j)$. We define a parallel layer of CPHASE gates in QFT as the following: 
\[
L_K = \{G(Q_i, ~Q_{i+1})\mid beg \leq i < ~end, step = 2\}.
\]

We also define a parallel layer of SWAP gates:
\[ S_K = \{ SWAP(Q_{i},~Q_{i+1}) \mid beg \leq i < end, step = 2\} \]

Then the LNN solution for QFT circuit (Fig. \ref{fig:LNN5}) can be described as repeated execution of \{$L_{K}$, $S_{K}$\}, and also using a loop below in Fig. \ref{fig:codesyn} (b).

\begin{figure}[h]
    \centering
    \includegraphics[width=0.45\textwidth]{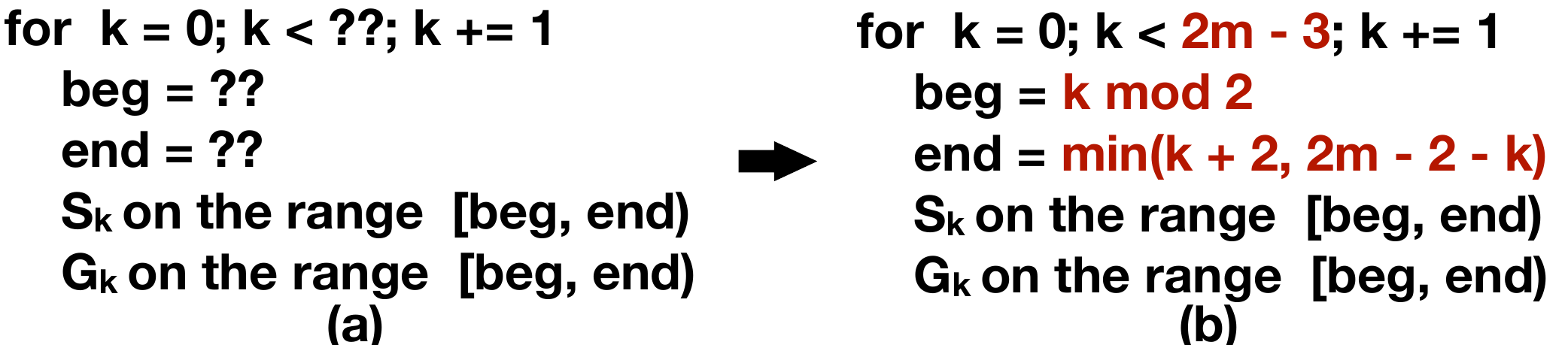}
    \caption{Code synthesis example}
    \label{fig:codesyn}
\end{figure}


In the code above, both $S_K$ and $G_K$ can be expressed as a loop too. For all the double-question-marks \textbf{``$??$"} in Fig. \ref{fig:codesyn} (a), if we replace them with proper functions solvable by program synthesis, we can find the QFT LNN mapping solution with ease. The challenge is that we have to make educated guesses about the shape of the code. But it is also an advantage since we can allow the integration of human intelligence (domain-specific knowledge and technical intuition) into the program synthesis process.

\subsection{Program Synthesis v.s. Prior Optimal Mappers}

 Prior studies for qubit mapping include the optimal SAT \cite{tan+:iccad20, tan+:iccad22, molavi+:micro22, murali+:asplos19} or A* \cite{zhang+:asplos21, jin+:asplos24} solvers. These mappers can automatically search the entire space of SWAP insertion, for instance, for each gate of the circuit, which SWAPs should be inserted before it can run. However, such search space is exponential. Moreover, these optimal solvers need to run every time when the QFT size changes.  
 
 These approaches do not explicitly explore the code structure of mapped QFT on regular architectures with arbitrary input sizes.     
  Zhang \etal \cite{zhang+:asplos21} find optimal solutions for QFT with less than 10 qubits, and then manually generalize patterns from the optimal solution of small circuits to larger size QFT circuits. But it does not scale well for complicated multi-dimensional architectures beyond 10 qubits. Assuming a QFT circuit on the N by N grid, if $N=4$, it requires at least 16 qubits, beyond the capacity of current optimal solvers. 

Our proposed approach has two-fold benefits:

\begin{itemize}
    \item  It allows users to embed certain structures (educated guesses) into the final solution of QFT mapping. These educated guesses represent human intelligence and domain-specific knowledge.
    \item It significantly reduces the search space. Since the final solution (functions to synthesize) needs to have a certain shape, it does not have to search in the original exponential space. 
\end{itemize}

\section{The Common Schemes}

We first propose a recursive QFT implementation by partitioning the qubits into subsets at the logical circuit level, which is the first step of how human intelligence is integrated into the program synthesis process.    

\subsection{Common Scheme 1: Recursive QFT Circuit}
\label{sec:recurQFT}

\textbf{\emph{2-partition QFT}}
First, we divide qubits into 2 subsets, and then divide the QFT process into three steps, without violating the dependence constraints. 

In this simple case, let's assume we have qubits $U = \{q_0, ..., q_{N-1}\}$. $U$ is divided into two sub-groups $U1$ (consecutive qubits $q_0, q_1, ..., q_k$) and $U2$ ( qubits $q_{k+1}, ..., q_{N-1}$). The sizes of $U1$ and $U2$ can be different. We can prove that the following steps are correct.

\begin{itemize}
    \item Step 1: Execute QFT on $U1$.
    \item Step 2: Interaction between $U1$ and $U2$, the relative order of gates is preserved.
    \item Step 3: Execute QFT on $U2$.
\end{itemize}

\vspace{-10pt}

\begin{figure}[htb]
    \centering
    \includegraphics[width=0.35\textwidth]{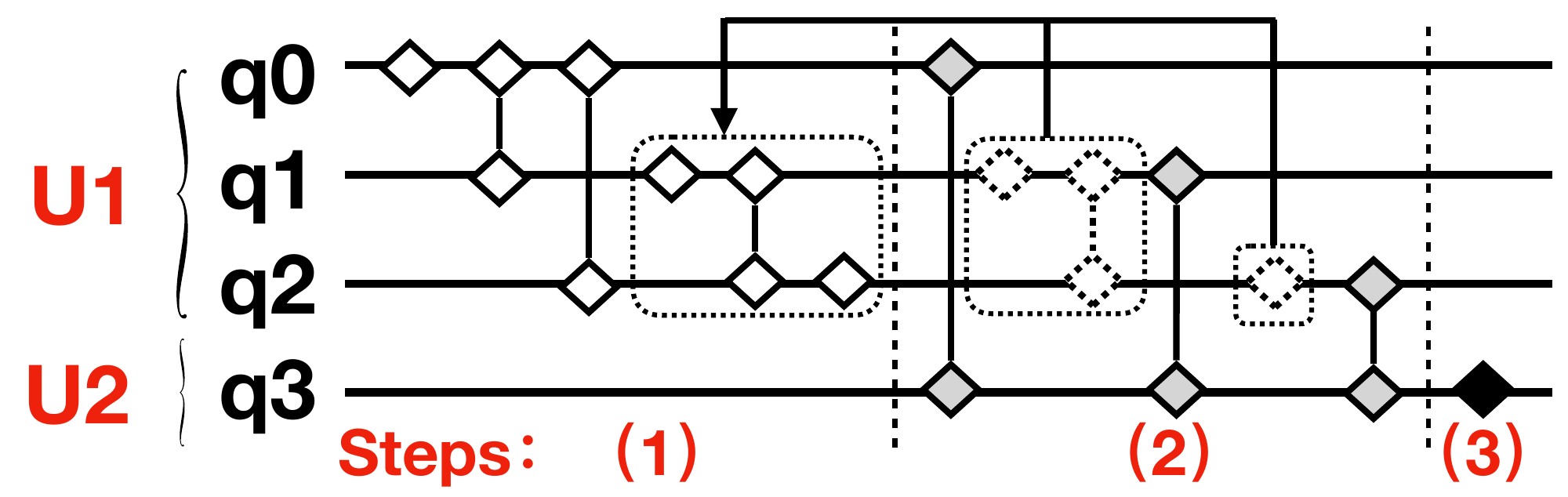}
    \caption{An example of our recursive QFT scheme for dividing 4 QFT qubits into two subsets $U1$ and $U2$. The dotted part shows the difference in the original QFT circuit.}
    \label{fig:qft_Divide&Conquer}

    \vspace{-12pt}
\end{figure}

Fig. \ref{fig:qft_Divide&Conquer} illustrates this idea. Due to the space limit, we sketch the proof for correctness. First, there are two types of dependence in QFT. 
\begin{itemize}
    \item \textbf{Type I dependence:} If two gates share the same control ( target), the one with a larger target (control) index must run after the one with a smaller index. For instance, $G1 = G(q_i, q_j)$ and $G2 = G(q_i, q_k)$, if $j < k$, G2 runs after G1. 
    \item \textbf{Type II dependence:} If one gate's control is another gate's target, the latter must run after the former. For instance, if $G1 = G(q_i, q_j)$, $G2=G(q_j, q_k)$, G1 must run before G2.
\end{itemize}

As long as a schedule satisfies these two types of dependence, it is valid. In the three steps above, we partition all gates involving qubits in $U1$ into two components: the interaction between themselves (Step 1), and the interaction between themselves and the rest of the qubits $U2$ (Step 2). Each component preserves its relative ordering before partition. We then first schedule all gates in Step 1 (within $U1$), all gates in Step 2 (between $U1$ and $U2$), and lastly Step 3 (all gates within $U2$).

For type I dependence, it is trivial to see this partition does not violate any of them. Just considering Step 1 and Step 2, for type II dependence, assuming $G1 = G(q_i, q_j)$, $G2=G(q_j, q_k)$, there are three cases for the location of $i$, $j$, and $k$. If all $i$, $j$, and $k$ are in $U1$, since we preserve the original relative order,  type II dependence is preserved. If $i$, $j$ are in $U1$, and $k$ in $U2$, G2 naturally runs after G1. If $i$ in $U1$, and $j$, $k$ in $U2$, it is not possible, because we are only looking at the interaction involving qubits in $U1$ in Step 1 and Step 2. G2 is not such a gate. The case that $i$, $j$, and $k$ are all in $U2$ is not possible, for the same reason. Step 3 is only for gates between qubits in $U3$ and naturally happens after all gates involving $U1$. Hence it is proved that such partition of QFT operations is correct.

\vspace{-5pt}

\begin{figure}[htb]
    \centering
    \includegraphics[width=0.4\textwidth]{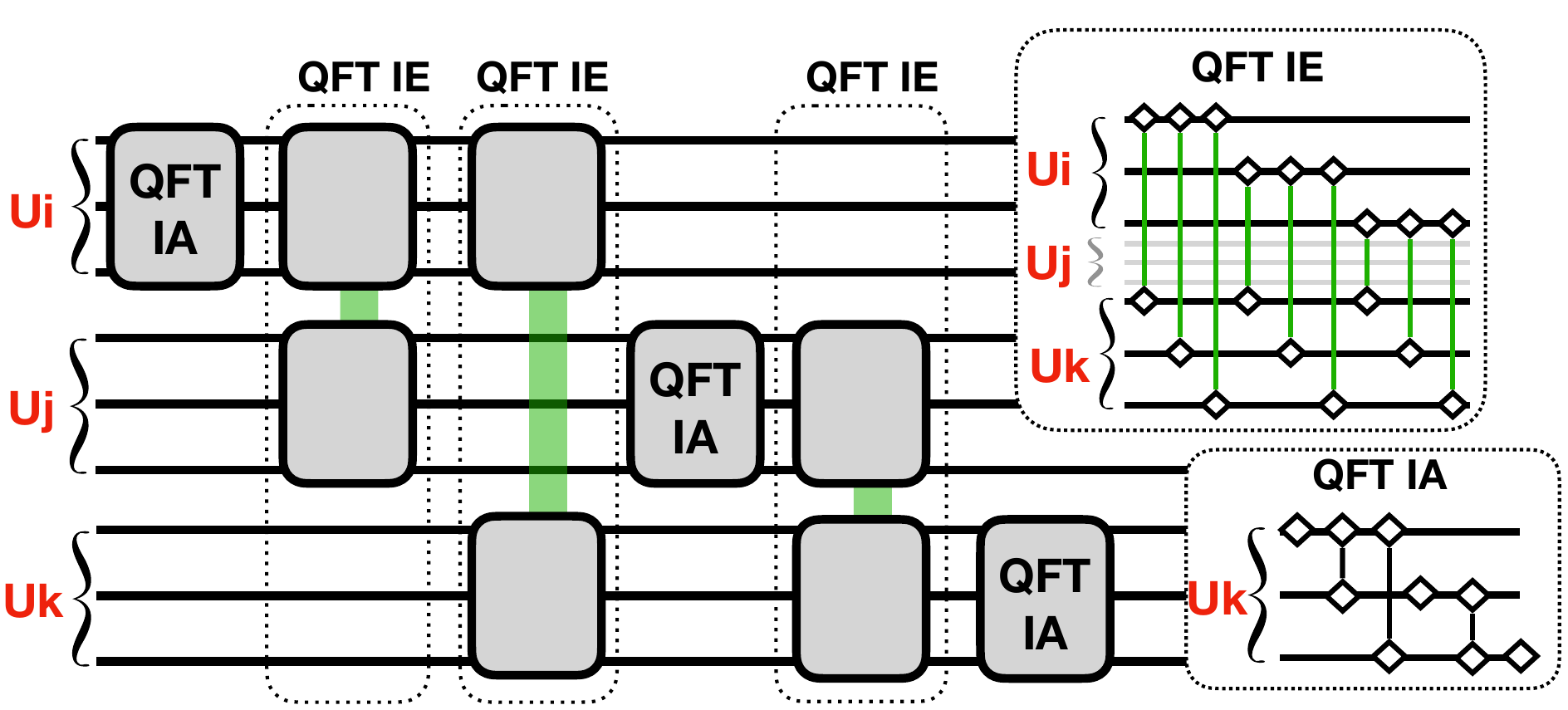}
    \caption{An example of the k-partition scheme for QFT. }
    \label{fig:qft_Div&Conq_K}
    
\vspace{-10pt}
\end{figure}

\textbf{\emph{$k$-partition QFT}}
In the previous example, we demonstrate the partition of QFT qubits into two sets, and the partition of QFT computation into three steps. It can in fact be extended to a k-partition QFT, where the qubits are divided into $k$ subsets $U_0, U_1, U_2, ...., U_{k-1}$. Again each subset needs not to have the same size. We can first divide the qubits into two subsets of $U_0$ and $U_{P2} = \{U_1, U_2, U_{k-1}\}$. Then we partition $U_{P2}$ into $U_2$ and $\{ U_3 ... U_{k-1}\}$. The whole process applies, and the proof of correctness still holds.

\begin{figure}[htb]
    \centering
    \includegraphics[width=0.4\textwidth]{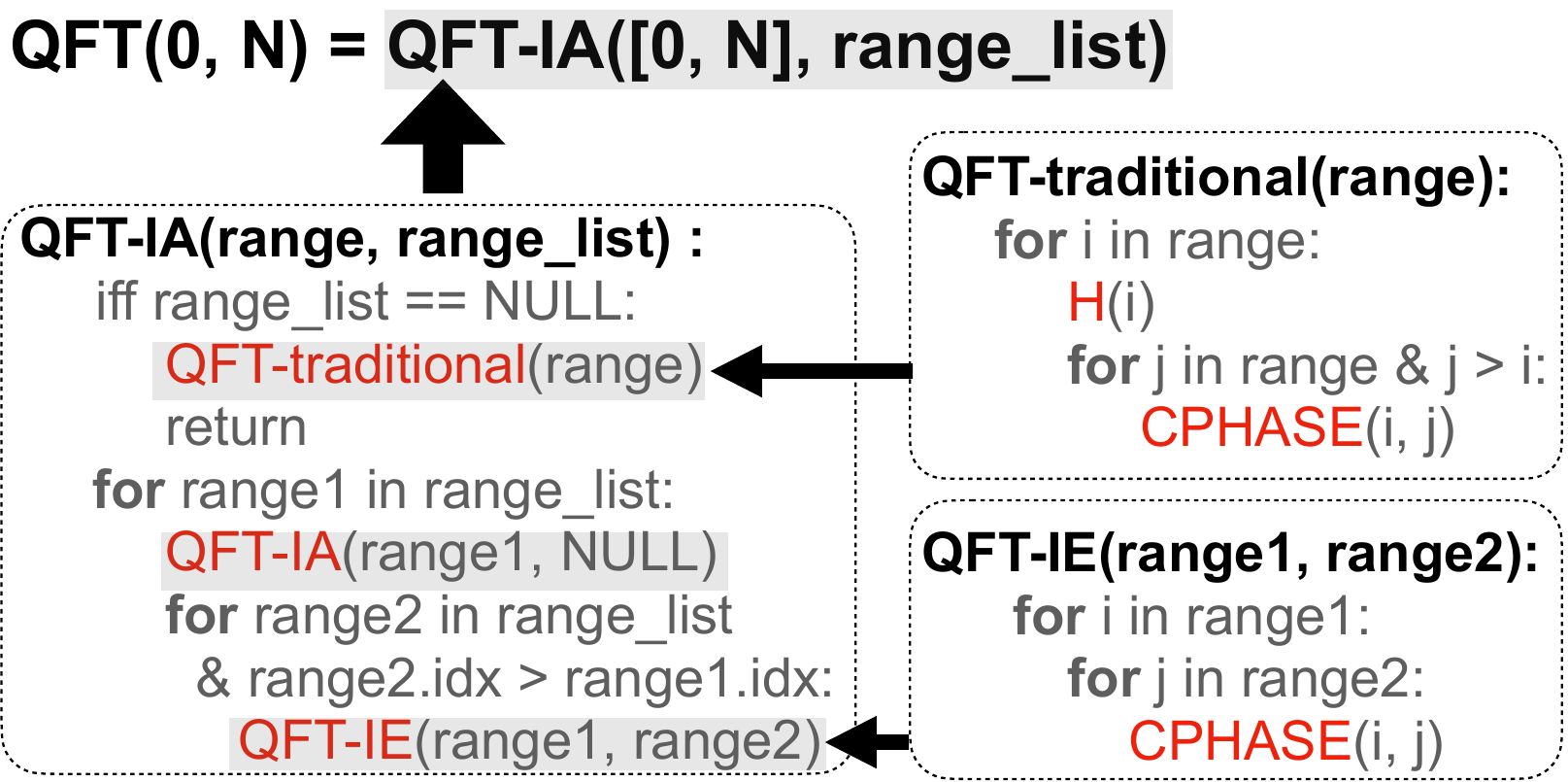}
    \caption{The k-partition QFT process. }
    \label{fig:DAC_Loop}
    \vspace{-20pt}
\end{figure}

We show the illustration of the k-partition QFT process in Fig. \ref{fig:qft_Div&Conq_K} and the pseudo-code in Fig. \ref{fig:DAC_Loop}. The function QFT-IA denotes QFT for a range of consecutive qubits and a list of small ranges.  The parameter $range\_list$ contains the list of small ranges. If the $range\_list$ is empty, QFT-IA degenerates to the traditional QFT operation, denoted as QFT-traditional, meaning we do not perform divide and conquer on this range of qubits. Otherwise, it performs a set of intra-QFT (QFT-IA) and  inter-QFT (QFT-IE) on and between $U_i$, where $0 \leq i < range\_list.size()$. QFT-IE allows two small ranges of qubits to interact using CPHASE gates, and these two small ranges can have different sizes.

\textbf{\emph{Discussion: Commutativity of CPHASE gates.}} Two CPHASE gates sharing one qubit can commute \cite{alam+:dac20, lao+:isca22}. This is most useful for QFT-IE since QFT-IE does not have any single-qubit gate, i.e., no Hardamard (H) gate. In QFT-IA, it is possible that two CPHASE gates sandwich a H gate, in which case, we cannot switch the order of the two CPHASE cases. QFT-IE operations can be flexible. 

We then have two different versions of QFT-IE. We denote the first version as \emph{QFT-IE-relaxed}, where gate reordering is exploited.  We refer to the second version as \emph{QFT-IE-strict}. Although QFT can directly use \emph{QFT-IE-relaxed}, we include the discussion for QFT-IE-strict mainly because there are other circuits with similar structure to QFT but do not use CPHASE gates for two-qubit interaction. Moreover, interestingly, we find that with strict ordering requirements on QFT-IE, the overall depth is still linear, and at most 2X of the version that allows reordering. This is a surprising finding in our paper.

\begin{figure*}[htb]
    \centering \includegraphics[width=0.63\textwidth]{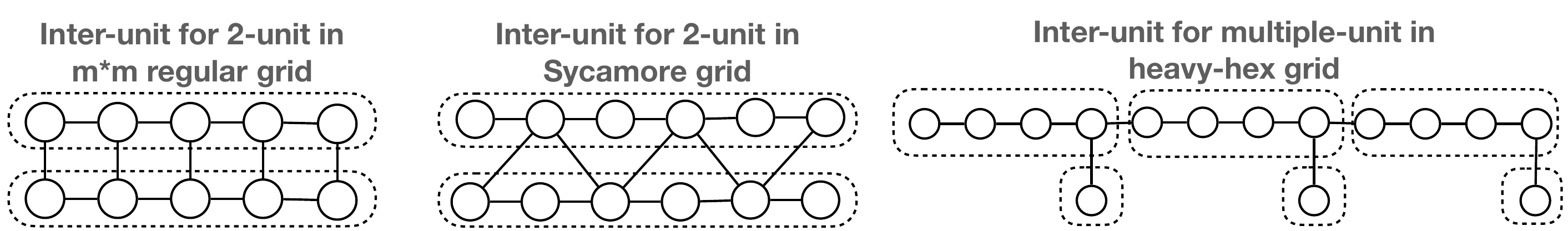}
    \caption{Unit partitions for different architectures, and the inter-unit case for each architecture. We unroll the heavy-hex architecture in the same way as that in \cite{weidenfeller+:arxiv22qaoa}. The longest path has one node hanging to it for every 4 nodes.  }
    \label{fig:unitpar}
\end{figure*}
\vspace{-0.1in}

\subsection{Common Scheme 2: Inter-unit QFT-IE Circuit}

Now it is established that we can divide a QFT circuit into multiple small QFT circuits and inter-unit QFT operations, by partitioning qubits into subsets. We can start describing our overall idea of QFT mapping on a complex architecture. 

 In the simple case, we can divide the qubits into equal-sized subsets. If each subset contains just one qubit, then it is equivalent to the original QFT circuit. If each subset contains more than one qubit, the whole process would require the broadly defined QFT-IE operations between two subsets of qubits, in Section \ref{sec:recurQFT} above. 

The recursive breakdown of QFT has two benefits. First, (1) we can reduce a multi-dimensional problem into multiple smaller-dimensional problems. Second, (2) the unit size is broadly defined and each unit can have a different number of qubits. This is important for the class of regular architectures which are difficult to be broken down into equal size units. 

For (1), recall that we can solve the LNN mapping problem. If the units can be connected as if they are on ``a line", one can run the unit-based linear QFT as shown in Fig. \ref{fig:unitQFTline}. The subtlety is that it needs to ``SWAP" two adjacent units on the ``line", and perform QFT-IE between two adjacent units on a ``line", both in a hardware-compliant manner. For Google Sycamore and 2D regular grid, the units can form a ``line". For instance, in a 2D regular grid, row 0 is connected to row 1, row 1 is connected to row 2, etc. 

For (2), we use it to hand the IBM heavy-hex architecture. It requires a more sophisticated higher-dimension divide-and-conquer method. We first reduce the overall problem into a simpler two-dimensional problem (as opposed to the one-dimensional problem in the decomposition step for Google Sycamore and 2D regular grid), and then we solve sub-problems. The sizes of the units are different in the IBM heavy hex case as shown in Fig. \ref{fig:unitpar}. 



\section{Linear-depth QFT on Sycamore}
\label{sec:sycamore}

\subsection{Unit definition}
\begin{figure}[htb]
    \centering
    \includegraphics[width=0.35\textwidth]{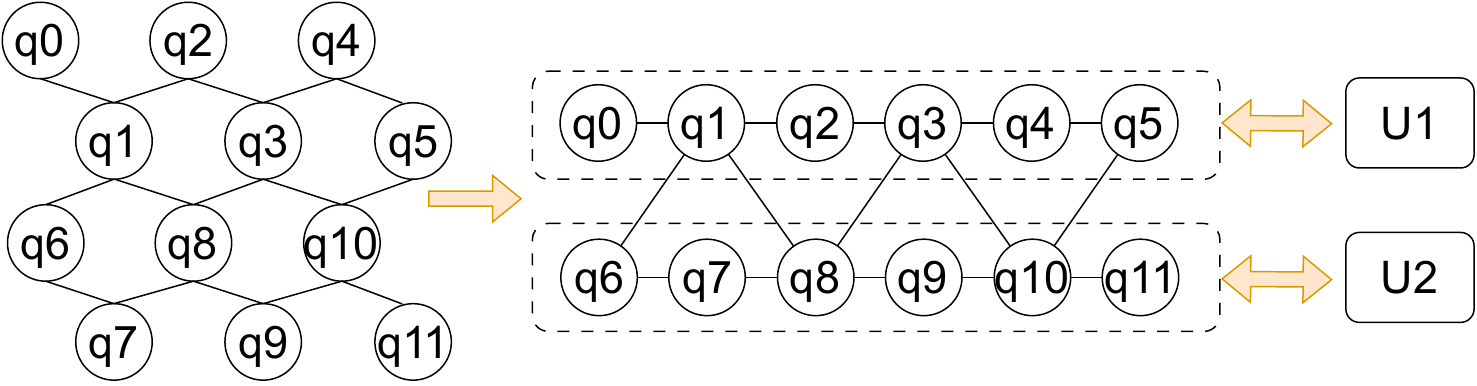}
    \caption{Unit definition in Sycamore architecture. }
    \label{fig:googlesycamore_unit}
    \vspace{-5pt}
\end{figure}

As is shown in Fig.~\ref{fig:googlesycamore_unit}, we combine every 2 rows together as a unit 
because we want to make sure that there is a path connecting all qubits within \textit{ONE} unit. Accordingly, as for $m * m$ Sycamore grid architecture, we could divide them into $m/2$ units, each of which contains $2*m$ qubits.

\vspace{-5pt}


Unit SWAP in this case is trivial. We describe it for the particular simple example in Fig. \ref{fig:googlesycamore_unit} above. 
{
\small
\begin{verbatim}
    parallelSWAP ({q1, q3, q5} {q6, q8, q10})
    parallelSWAP ({q1, q3, q5} {q7, q9, q11}) 
                 ({q0, q2, q4} {q6, q8, q10})
    parallelSWAP ({q0, q2, q4} {q7, q9, q11})
\end{verbatim}
}
This design makes it possible to complete the unit swap between 2 adjacent units in 3 steps. This can be generalized.  


\subsection{Intra and inter-unit QFT}
We decompose the qubit mapping problem of QFT over the Sycamore architecture into 2 categories: intra-unit and inter-unit mapping.

\textbf{\emph{Intra-unit QFT}} Given all qubits within one unit are in a line, we leverage the mapping algorithm used for QFT over LNN architecture for intra-unit qubit mapping.

\textbf{\emph{Intuition for inter-unit QFT}} This is the most interesting part for Sycamore architecture. 
Before finding a reasonable algorithm to do inter-unit QFT over Sycamore architecture, we want to show something that inspired us. It is the existing solution of LNN QFT. This idea is illustrated in Fig. \ref{fig:LNN5}. It is only for qubits on a single line, but it serves as a good starting point for us to explore the 2-unit interaction problem in the Sycamore architecture.

In this LNN solution, each layer (a layer is a set of gates running in parallel) consists of consecutive disjoint SWAPs starting from either qubit 0 or 1, and ending at a location that is somewhat a linear function of the time step. Each qubit travels along a path. For any pair of qubits, their paths will cross at some point. Right after (or before) 2 qubits' paths cross, they are of distance 1. This enables a CPHASE gate operation between two qubits. Moreover, any two qubits meet once and only once. This is important. 

It can be seen that in Fig. \ref{fig:LNN5} each qubit would move towards the physical location ``Q0", then stop there for a time step, change the direction, and keep moving before reaching its destination. Each qubit's destination is the mirrored location of its original location; for instance, qubit $q0$ starts at location Q0 and then ends at location $5$, qubit $q1$ starts at location $Q1$ and then stops at location $Q4$.

\begin{figure}[htb]
    \centering
    \includegraphics[width=0.3\textwidth]{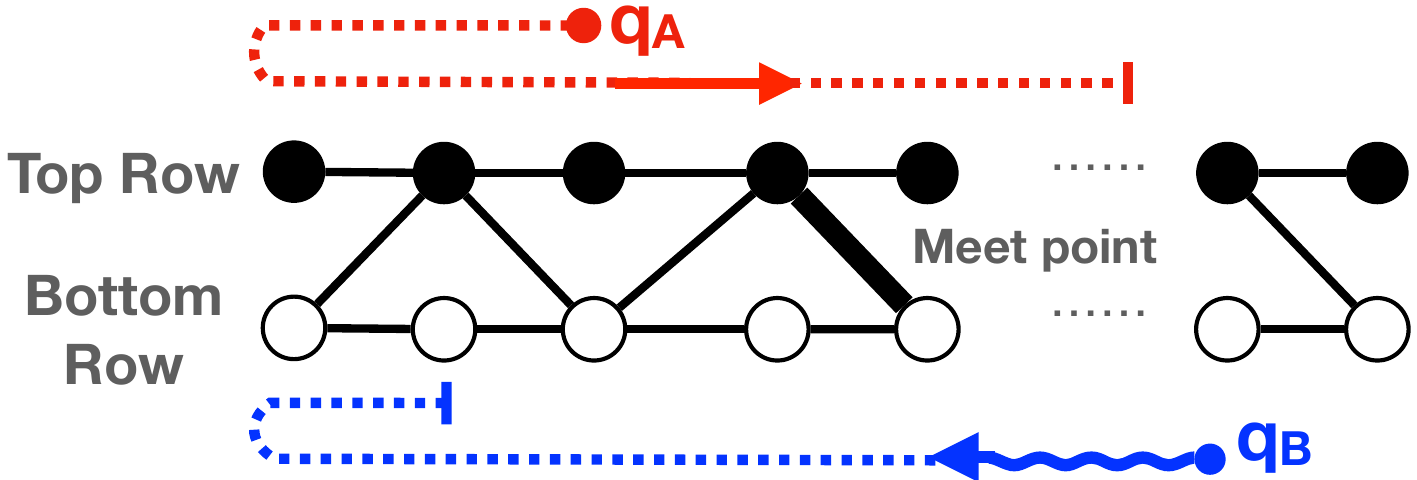}
    \caption{Condition for 2 qubits in two different units to meet in Google Sycamore. }
    \label{fig:sycamorepathcross}
\end{figure}

In the two-unit case, it is as if there are two lines. We refer to them as the top row and the bottom row as shown in Fig. \ref{fig:sycamorepathcross}. If a qubit in the top row moves along a path, and a qubit in the bottom row moves along a path, and somehow they meet through a link between these two rows, then we can run a CPHASE gate. The uniqueness in Sycamore's inter-unit case is that there is a connection between one qubit on the top row and the qubit on the bottom row \emph{if the two qubits' column index differ by 1}, as shown in Fig. 
\ref{fig:sycamorepathcross}.  

The challenge is how to move qubits in the top row and bottom row, and let them meet. Our guess is that we may sync the movement of qubits on the top row and bottom row. The reason is that if qubits are on a single line, any two qubits will have a distance of 1 at some time point, if they follow the movement pattern in LNN. This is analog to the case that the column index differs by 1 for two linked qubits from the top row and bottom row respectively.

\begin{figure*}[htb]
    \centering
    \includegraphics[width=0.8\textwidth]{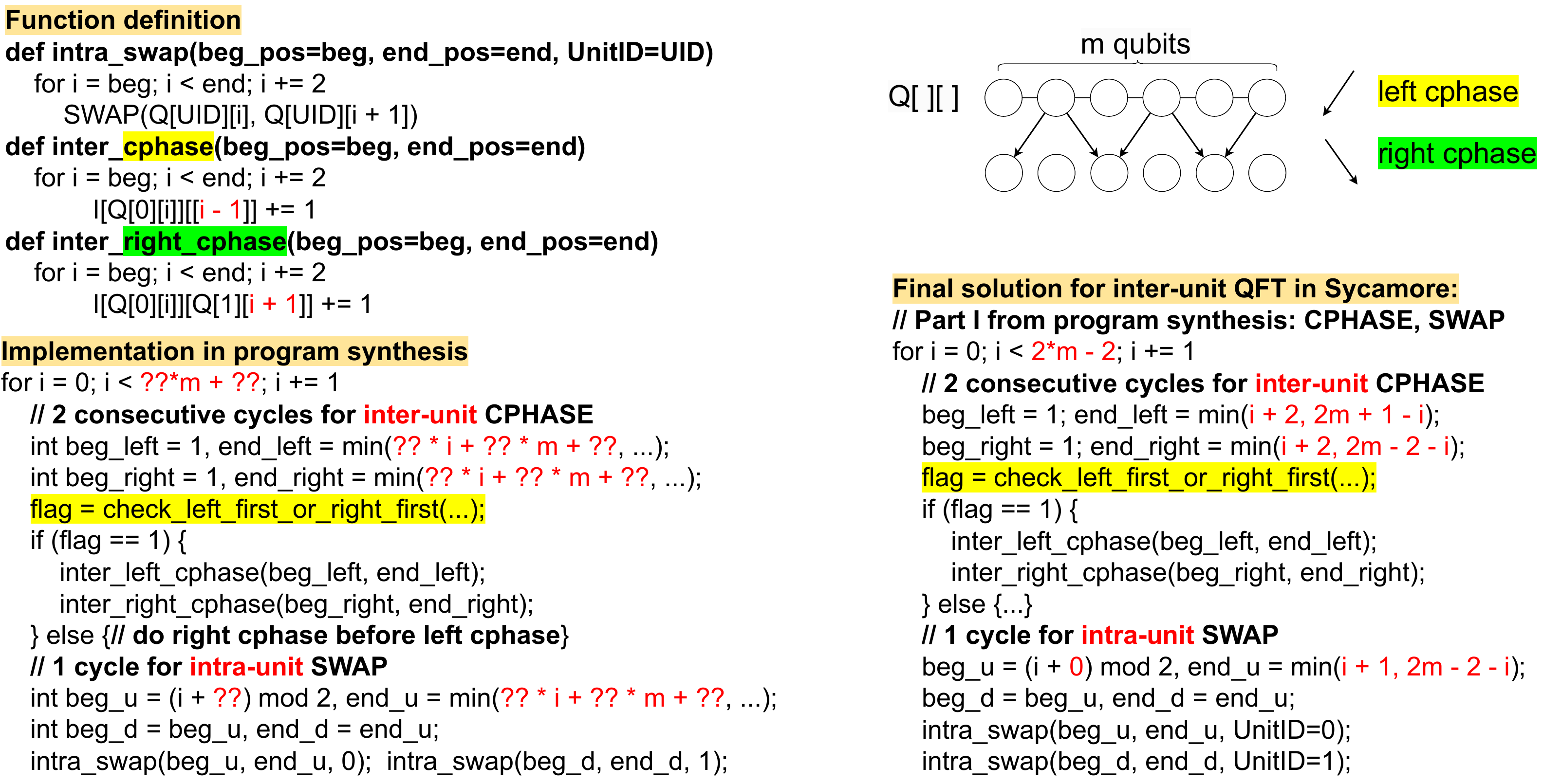}
    \vspace{-0.1in}
    \caption{Program synthesis implementation and outcome for Sycamore.}
    \label{fig:impl_out}
    \vspace{-0.1in}
\end{figure*}

But one also has to be careful, in the Sycamore architecture, two vertices having the same column index on the top and bottom row, do not have a direct connection between them. Moreover, between two units, there are $row\_size - 1$ links.  Fortunately, it is simple to figure out a solution for CPHASE gates between pairs of qubits in the same column. We can SWAP one of them horizontally with its neighbor at the top (bottom) row, keep the other one on the bottom (top) row unchanged, run a CPHASE gate, and then revert the qubit back to its original location with the same SWAP.

Hence, we really want to ask one question: \textit{if we exclude all CPHASE gate operations for qubit pairs in the same column initially, could program synthesis solve the rest gate operations assuming we sync the steps of the top and bottom row?} 
If there is such a solution that accords with our assumption, it might be possible for us to generalize something useful from it.


\vspace{-10pt}

\subsection{Program Synthesis for This Architecture}

\paragraph{Specification}  
First of all, we enable the program synthesis tool to exclude CPHASE gate operations between qubits that are in the same vertical lines initially (e.g., in Fig.~\ref{fig:googlesycamore_unit} all ($q_i$, $q_{i + 6}$) pairs). The excluded gate operations will be fixed using the way mentioned above.

Additionally, we allow a CPHASE gate to happen more than once between two qubits just for synthesis purposes. This is to account for the potential boundary case in a loop, where there might be a head and tail condition that is different from the repeated steps in the middle. The head and tail conditions are boundary conditions.  Later, if the program synthesis engine successfully figures out the repeated steps, we can fix the solution by manually setting the head and tail components of the loop, and ensuring only one CPHASE between each pair of qubits.  



\textbf{\emph{Implementation}}  The implementation is shown in Fig. \ref{fig:impl_out}. The intuition is multi-fold. First, we maximize the utilization of all links between 2 units. We run all CPHASE gates between two units as long as their dependence is resolved. Since the links between two units form a line, it needs at least two steps to run all of them (or a consecutive subset of them), we divide them into the left and right CPHASE gates.

Second, the starting point for the SWAPs in one loop iteration should be different from that in the next loop iteration. Otherwise, it means nothing but to undo the previous SWAP; therefore, we use the ``mod" function to reflect this feature. 
\begin{figure*}[htb]
    \centering
    \includegraphics[width=0.95\textwidth]{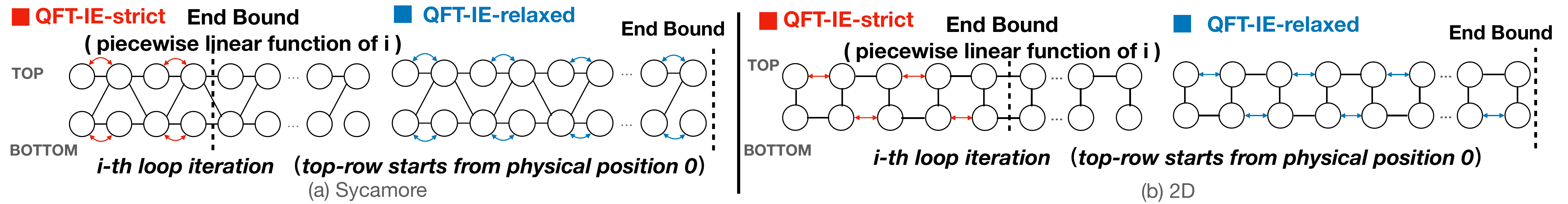}
    \vspace{-10pt}
    \caption{QFT-IE-strict and QFT-IE-relaxed in Sycamore and 2D grid. Note that the top row starts at position 0 in this example, but it in fact alternates between 0 and 1 in consecutive loop iterations for both Sycamore and 2D grid.  We omit the 1 case due to the space limit. (a) Sycamore, and (b) 2D grid.  }
    \label{fig:QFT-IE-Sycamore-strict-relaxed}
\end{figure*}

Third, inspired by the SWAP bound in the LNN case in Fig. \ref{fig:LNN5}, where a triangle shape covers the set of valid SWAP steps.  We ``guess" that the bounding point for the SWAPs in one layer might follow a piecewise linear function of loop induction variable represented using the ``min(...)" function. The \emph{min(...)} function may take more than 2 arguments, but we find two are enough in our case. 

\textbf{\emph{Discussion}} So far, we have discussed the QFT-IE-strict case, where the type 1 and type 2 dependencies must be satisfied. Surprisingly we were also able to find a solution for the QFT-IE-relaxed case using the same code shape in the synthesis process. The only difference is that we relaxed the dependence constraints for the QFT-IE-relaxed case. We discuss our discovered solution for both below.

\subsection{QFT-IE-strict and QFT-IE-relaxed Solutions}
The inter-unit QFT-IE-strict solution is presented in Fig. \ref{fig:impl_out} in pseudocode. The repeated steps are visualized in Fig. \ref{fig:QFT-IE-Sycamore-strict-relaxed}. 

The inter-unit QFT-IE-relax solution is below:
\begin{small}
\begin{verbatim}
    for (i = 0; i <= m; i += 1)
       CPHASE on all inter-unit connections   
      // Intra-unit swap
      beg = i mod 2
      intra_swap(beg_pos=beg, end_pos=m, UnitID=0)
      intra_swap(beg_pos=beg, end_pos=m, UnitID=1)
\end{verbatim}
\end{small}

The repeated steps are shown in Fig. \ref{fig:QFT-IE-Sycamore-strict-relaxed} (a). In the implementation of the piecewise linear functions, the QFT-IE-relaxed version in fact chooses to use the same function for two parameters in min(). But QFT-IE-strict chooses two different linear functions. The QFT-IE-relaxed version is two times faster than the QFT-IE-strict version.

One extra benefit of our solution is that both QFT-IE-strict and QFT-IE-relax will mirror the position of all qubits within a unit. Hence it facilitates further processing that for QFT-IA. In Google Sycamore, since QFT-IA uses the LNN solution, the qubits must be placed in natural number ordering in the LNN, or the reversed natural number ordering.


\subsection{Time Complexity}

\begin{figure}[htb]
    \centering
    \includegraphics[width=0.5\textwidth]{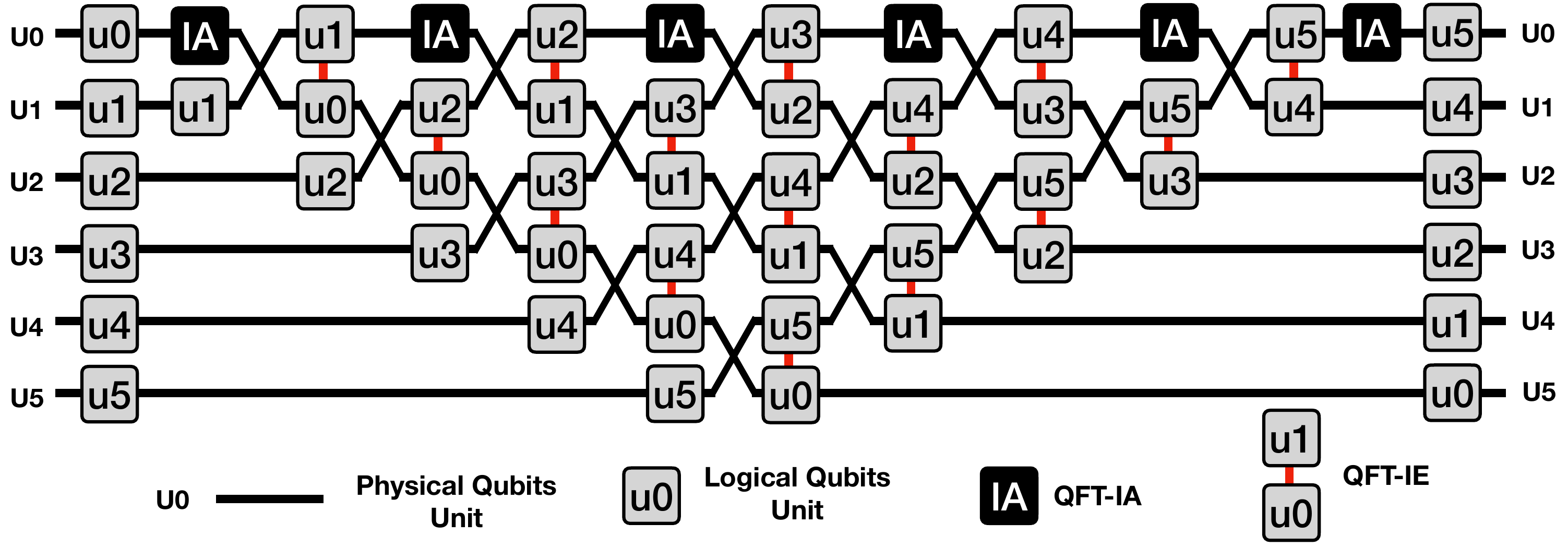}
    \caption{Unit-wise QFT using the recursive QFT scheme. This is analogous to the original LNN QFT solution. }
    \label{fig:unitQFTline}
\end{figure}

\paragraph{Complexity over m*m Sycamore} 
Assuming $N=m*m$ Sycamore grid, where m is the original row size.  In our formulation, each unit consists of 2m qubits and there are $\frac{m}{2}$ units. Each unit is connected as if they are on a line, as described earlier. We will do the QFT-IA and QFT-IE using the divide-and-conquer method. The hardware-compliant unit QFT on LNN is presented in Fig. \ref{fig:unitQFTline}. 

\textbf{\emph{QFT-IE-relaxed}}
For the QFT-IE-relaxed case, each QFT-IE takes $3*(2m+1)$ time steps. Each QFT-IA takes $4*(2m)-6$ time steps. 
Each unit SWAP takes 3 time steps. There are in total $(m/2)+O(1)$ QFT-IE parallel steps, and in total $(m/2)+O(1)$ (QFT-IE, QFT-IA) mixed steps. There are total $2*(m/2)-3$ unit SWAP steps. Hence, the total time complexity is $7m^2+O(m) = 7N + O(\sqrt(N))$, where N is the total number of qubits.

\textbf{QFT-IE-strict} For the strict ordering, QFT-IE will have the same complexity as QFT-IA in our solution. Hence the overall complexity is $8N+O(\sqrt{N})$.

\section{Linear-depth QFT on 2D Grid}

\subsection{Unit Definition}

 We consider each row in the 2D grid as a unit and place qubits in natural number ordering from left to right, and from top row to bottom row. This is a row-major order initial layout. Similar to Google Sycamore, the problem could be solved by tackling three sub-problems: intra-unit interaction QFT-IA, inter-unit interaction QFT-IE, and unit swapping.

The unit swap is trivial by applying transversal SWAPs using the vertical links between two units in one step.
\vspace{-5pt}

\subsection{Intra-unit and inter-unit QFT}

\paragraph{Intra-unit QFT}
The structure of a single unit is exactly the same as the LNN structure. We leverage the existing solution for QFT on LNN for QFT-IA.

\textbf{The intuition for inter-Unit QFT}
The intuition is contrary to that of the inter-unit in Google Sycamore. Recall that we ``guessed" in Google Sycamore, that the top row and bottom should move in sync, but now this should not work.   

\begin{figure}[htb]
    \centering
    \includegraphics[width=0.3\textwidth]{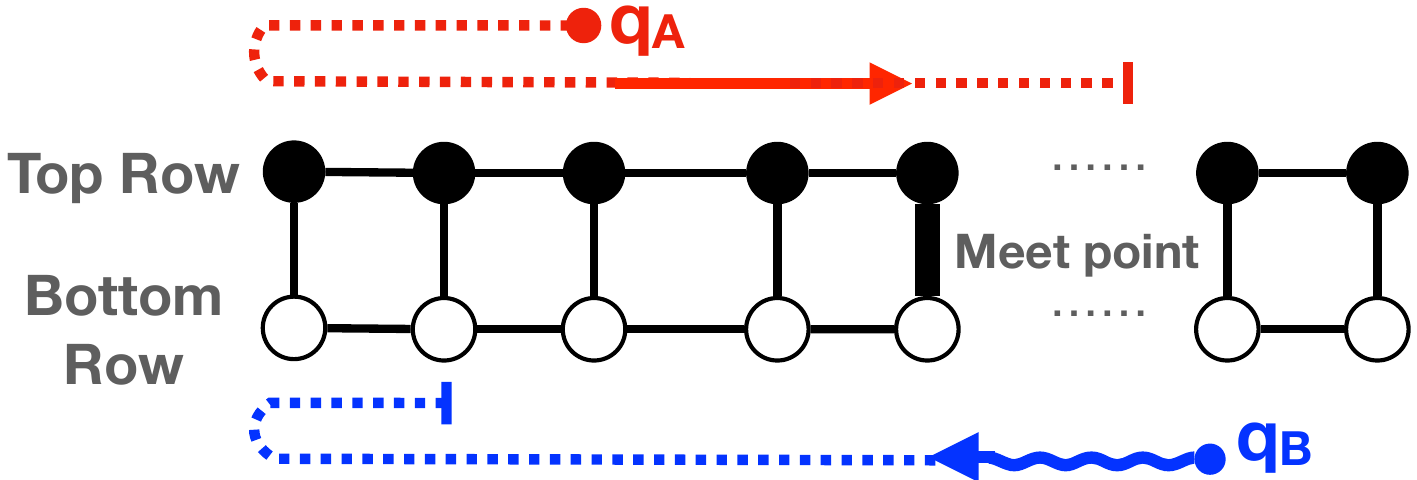}
    \caption{Trajectory of two points from the top row and the bottom row.}
    \label{fig:pathcross}
\end{figure}

If we look at path crossing in Fig. \ref{fig:pathcross}, now the two qubits on the top row and bottom row should meet at exactly the same column, as opposed to two columns of distance 1. This is because a 2D grid has direct vertical links between qubits in the top row and the bottom row. 

In this case, the top-row qubits and bottom-row qubits should not move in the same way. Otherwise, for any column at any time instant, the top qubit's neighbor in the bottom row is always the same. This makes it impossible to do all-to-all interaction between these two rows.

Now our guess is that each qubit should still travel along a path, but there must be a difference between the starting points of movement (SWAPs) in the top row and the bottom row. Two qubits on different rows will have their paths cross at some point (meeting in the same column).  This leads to our program synthesis model as follows.

\subsection{Program Synthesis for This Architecture}

\paragraph{Specification}

The specification is similar to that in Google Sycamore. The only difference is that we do not exclude the qubit pairs in the vertical links at initial qubit mapping.

\begin{figure}[htb]
    \centering
\includegraphics[width=0.45\textwidth]{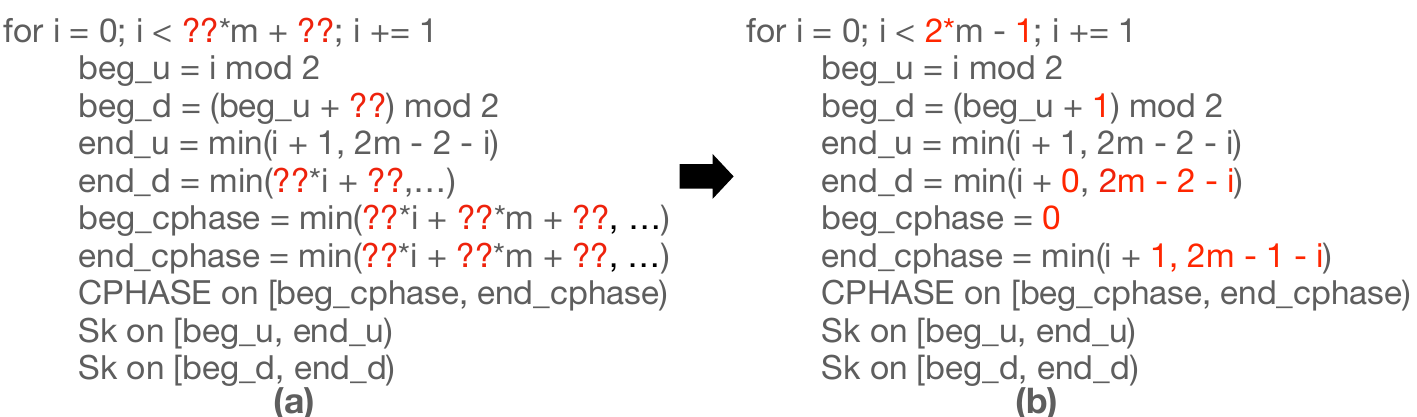}
    \caption{Synthesized code for inter-unit of the 2D regular grid in the strict version. }
    \label{fig:syn2dIE}
\end{figure}

\textbf{Implementation.} With the above conditions, our formulated code shape is in Fig. \ref{fig:syn2dIE}.  

In the code in Fig. \ref{fig:syn2dIE}, ``beg\_u" and ``end\_u" respectively represent the begin and end location for the SWAPs in the top row. ``beg\_d" and ``end\_d" respectively represent the begin and end locations for the SWAPs in the bottom row. CPHASE gates operate on a range of vertical links, and the range is a linear function of the loop induction and loop invariant variables too. $m$ is the unit size.

We specifically add a difference between the starting location of the top-row SWAP, and that of the bottom-row SWAP, using the modular function ``beg\_u + ??" mod 2.  We simplified the loop by fixing the movement of the top-row as that in QFT LNN. In this case, the simple code shape actually worked. It found a solution successfully, shown in Fig. \ref{fig:syn2dIE} (b). 

\begin{figure}[htb]
    \centering
    \vspace{-7pt} \includegraphics[width=0.47\textwidth]{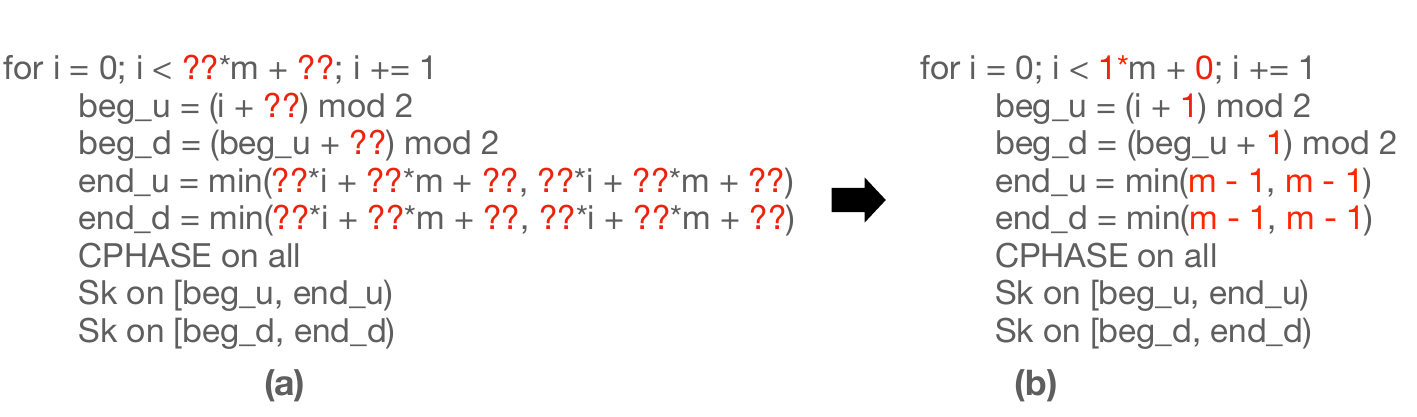}
    \vspace{-5pt}
    \caption{2-unit interaction for the relaxed order QFT-IE.}
    \label{fig:syn2drelaxed}
    \vspace{-0.1in}
\end{figure}

We took a step further and chose not to fix the movement in the top row. Rather, we use piecewise linear functions to specify the beginning and ending locations for intra-row SWAP, as shown in Fig. \ref{fig:syn2drelaxed}. Using this approach, we found a solution for QFT-IE-relaxed as shown in Fig. \ref{fig:syn2drelaxed} (b). The repeated step is visualized in Fig. \ref{fig:QFT-IE-Sycamore-strict-relaxed} (b).

\vspace{-0.1in}
\subsection{Time Complexity}

As for the time complexity, we use unit QFT on a line, as shown in Fig. \ref{fig:unitQFTline}. Assuming we have $N=m*m$ qubits. Each unit has $m$ qubits. 

\textbf{\emph{QFT-IE-Relaxed}} Each QFT-IA has complexity $4m-6$. Each QFT-IE has complexity $2m$. Each unit SWAP takes one time step. There are $m + O(1)$ QFT-IE steps. There are $m+O(1)$ mixed (QFT-IA, QFT-IE) steps. In total, the complexity is $6m^2 + O(m) = 6N + O(\sqrt{N})$.

\textbf{\emph{QFT-IE-Strict}} Since the strict version doubles the time complexity compared with the relaxed version. The overall complexity is $8N + O(\sqrt{N})$

\section{Heavy-hex}
\label{sec:heavyhex}

\begin{figure}[htb]
    \centering
    \includegraphics[width=0.49\textwidth]{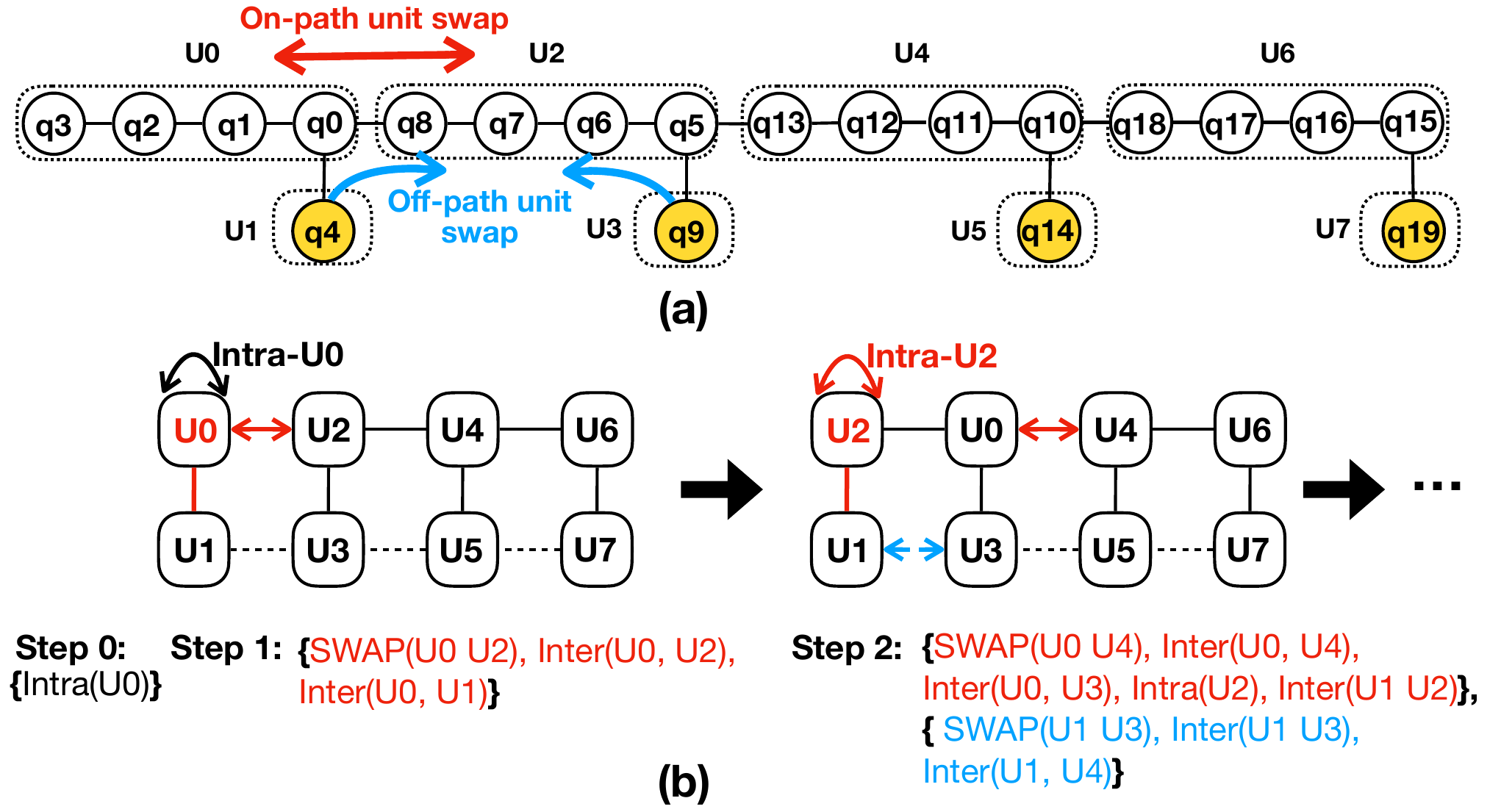}
    \caption{Heavy-hex in 2D Layout with Column Major. (a) 20-qubits Heavy-hex architecture with qubits grouping. (b) Details of each step by following the pattern of implementing QFT in 2D grid architecture.}
    \label{fig:heavy_hex_to_unit_col}
    \vspace{-10pt}
\end{figure}

\textbf{\emph{Unit Definition: }} In a heavy-hex architecture, the unit definition is different from other architectures due to its sparse and relatively irregular structure. In heavvy-hex architecture, we define two types of units, \emph{on-path unit} and \emph{off-path unit}. One on-path unit contains four qubits from the longest path and one off-path unit consists of one qubit that hangs to the longest path. For example, qubit q0-q3 form one on-path unit in Fig. \ref{fig:heavy_hex_to_unit_col} (a), and an off-path unit U1 is hanging to the first on-path unit U0 with qubit q4. By grouping qubits in this way, we can imagine that the connection between units is analogue to the 2D grid architecture, as the result shown in Fig. \ref{fig:heavy_hex_to_unit_col} (b). However the 2D grid placement of units is not row-major, while the previous solution we have for 2D grid inter-unit is row-major. Note this is a key difference. 

\begin{figure}[htb]
    \centering
    \includegraphics[width=0.4\textwidth]{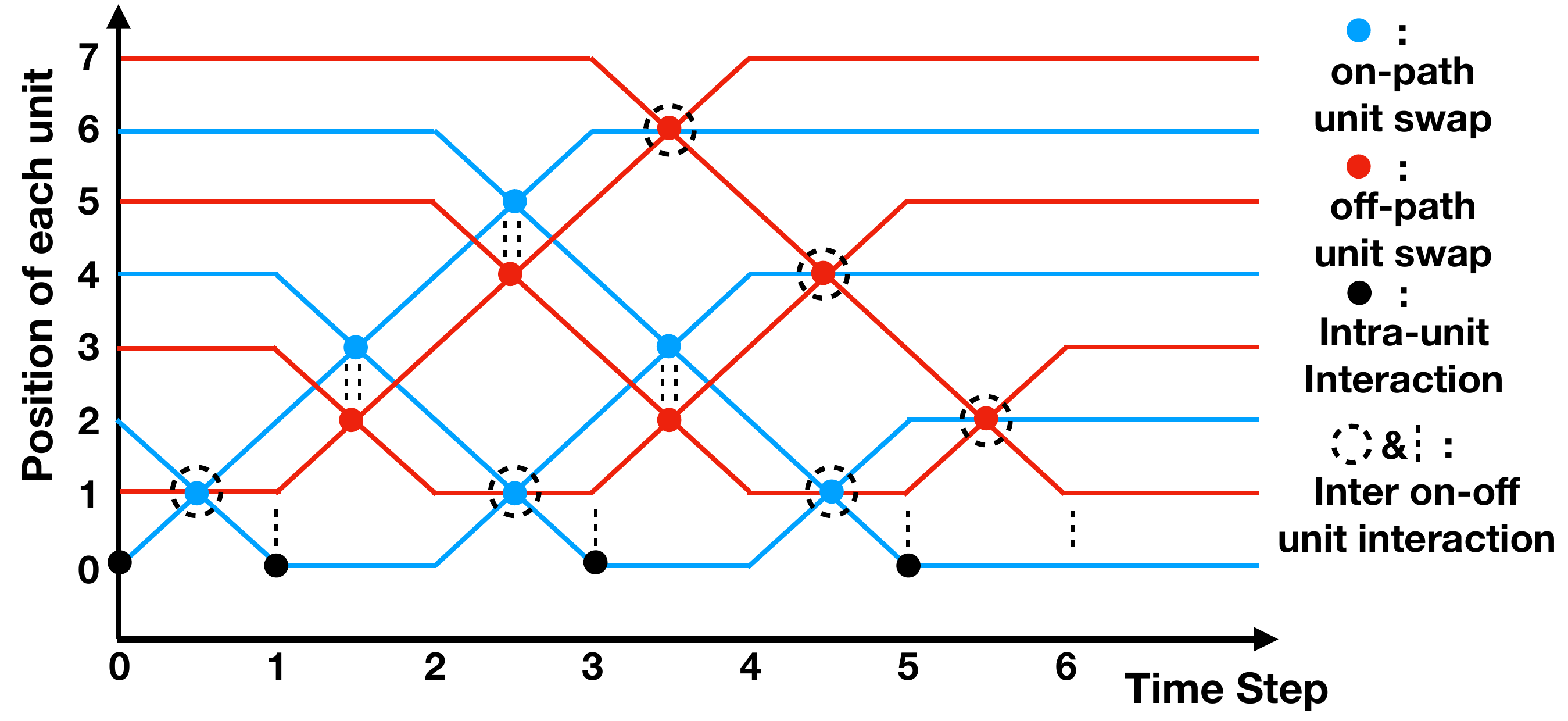}
    \vspace{-0.1in}
    \caption{The trajectory of each unit in the unit-wise 2D grid. The initialing mapping of all units in alphabetical order. }
    \vspace{-0.1in}
    \label{fig:hex_time_position}
\end{figure}

The connections between top units are real physical connections, and bottom units are connected by virtual connections.   The initial mapping for each logical qubit is determined in these units' construction in Fig. \ref{fig:hex_time_position} in order to facilitate our next processing step for optimizations. We discuss the case in Sec. \ref{sec:fh} for real IBM devices. 

\textbf{\emph{Unit Routing and Interactions}}
Since we represent the heavy-hex architecture by a 2D (2xM) grid in column-major order, we can follow the 2xM QFT pattern for unit movements and interactions, introduced in \cite{zhang+:asplos21}. In such a 2xM QFT pattern, there are three major operations: intra-row interactions, intra-row swapping, and inter-row interactions.  In our case, one intra-row interaction contains intra-on-path units interactions and intra-off-path units interactions; one intra-row swapping contains on-path units swapping and off-path units swapping;  one inter-row interaction stands for one inter-on\&off-path units interaction.

 We represent that pattern by a position-time diagram in Fig. \ref{fig:hex_time_position}. The blue line represents the trajectory of units on the top row, and the red line represents the trajectory of units on the bottom row. In this figure, we can see all the details in  2xM QFT unit-wise pattern. For example, the top row units are one more step faster than the bottom row qubits. The cross of two lines from the same color stands for a unit swap. Note that a SWAP on the top row could happen in parallel with a SWAP in the bottom row in a 2D grid pattern. In heavy-hex, those swaps happen asynchronously, since the off-path unit swap needs assistance from the on-path unit. 
 

\textbf{\emph{Solving Sub-problems:} }
\emph{Solving on-path unit interaction} reuses the divide-and-conquer scheme shown in Fig. \ref{fig:qft_Div&Conq_K}. The on-path qubits are divided into a sequence of on-path units. The on-path unit interaction is also divided into three sub-problems, intra-on-path-unit interaction, inter-on-path-unit interaction, on-path-unit swapping. Solving the intra-on-path-unit interaction could simply reuse the solution for compiling QFT in LNN architecture, since the qubit connection for an on-path unit is also LNN. 

Next, the on-path-unit swapping and inter-on-path-unit interaction could happen simultaneously. In Fig. \ref{fig:heavy_hex_to_unit_col}(a), U0 and U1 are connected in an LNN with 8 qubits. Moving U0 and U1 toward each other would make each qubit in U0 adjacent to each qubit in U1. 

\emph{Inter-unit interaction for off-path unit} has to rely on the on-path unit in between of them. Since there are no real links between off-path units. As the example in Fig. \ref{fig:heavy_hex_to_unit_col}(a), we can swap U1 and U3 with the help of U2 and implement the QFT-IE(U1, U2).

\emph{Inter-unit interaction between on-path unit and off-path unit } could occur at three procedures. Firstly, it happens when on-path-unit swapping happen. For example, as the qubits highlighted in Fig. \ref{fig:heavy_hex_to_unit_col}(a), qubits q0-q4 all pass through the right-most position in unit U0 and adjacent to the neighboring unit U1 once. Secondly, it could also happen when off-path-unit swapping happens. In Fig. \ref{fig:heavy_hex_to_unit_col}(a), if we want to swap U1 and U3, qubit q4 will travel through U2 and allow us to do Inter(U1, U2). Lastly, when an on-path does the intra-unit operations, we can do the inter-unit interaction between on-path unit and off-path unit simultaneously. For example, Intra(U0) would make each qubit in U0 adjacent to U1 once.  To make qubits from the on-path unit interact with the off-path unit, qubits within an on-path unit have to be mapped in descending order.

\begin{figure}[htb]
    \centering
    \includegraphics[width=0.4\textwidth]{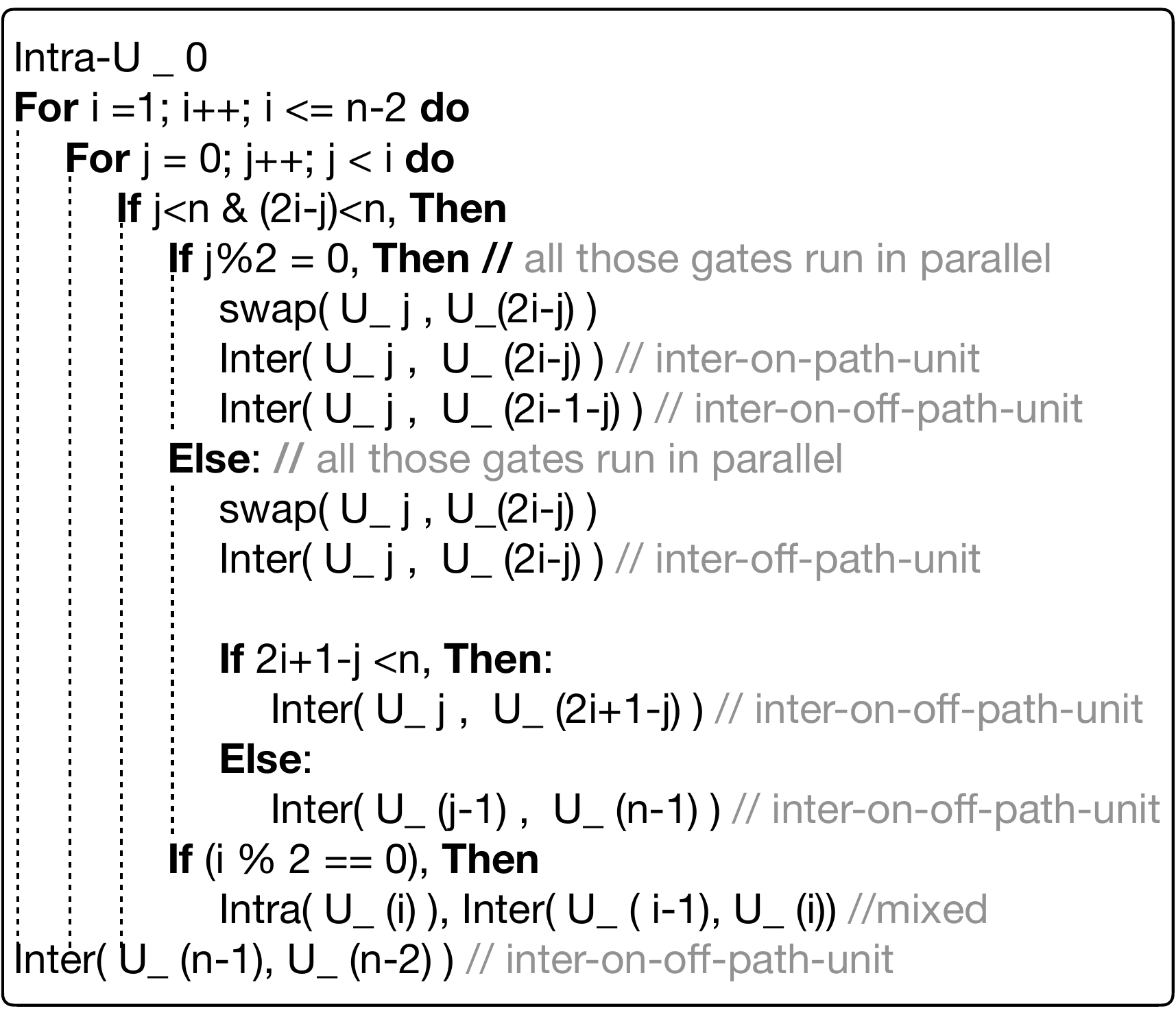}
    \caption{Loop of Implementing QFT in Heavy-hex. }
    \label{fig:hexloop}
    \vspace{-10pt}
\end{figure}

\textbf{\emph{Time Complexity}} As is shown in Fig. \ref{fig:heavy_hex_to_unit_col}, each on-path units swapping contains the unit swapping operations, inter-on-path unit interaction, inter-on\&off-path interaction. The cost of one mixed operation is 18 steps. 

Meanwhile, intra-on-path unit interaction and the inter-on\&-off-path unit interaction occurred by it could be implemented simultaneously with a mixed operation. 

For example, in Fig. \ref{fig:heavy_hex_to_unit_col}(b) step2, intra(U2), inter(U1, U2) could run simultaneously with a mixed operation which includes SWAP(U0, U4), Inter(U0, U4), and Inter(U0, U3). In this new layout, we also have off-path units swapping. This step is implemented after its corresponding on-path units swapping. For instance, SWAP(U1, U3) happens after SWAP(U0 U4).  The cost of one on-path unit swapping is 18 steps.  The cost of one off-path units swapping is 11 steps. In total, there are 29 steps in one unit swapping step. For a 2D grid layout, each row has N/5 units, then, such unit swapping happens (N/5-2)*2 times, N is the number of qubits. So the overall time complexity is 11.6N+O(1).

\section{Fault Handling}
\label{sec:fh}

\textbf{\emph{Handling the error in Heavy-hex.}} For qubits that have high error rates, we will exclude them from the computation. We refer to them as faulty qubits. Errors happen at the on-path unit or off-path unit.  Errors that occur at the on-path unit might isolate the whole architecture into two disconnected pieces. QFT cannot be implemented between such disconnected components.

We handle faulty qubits that occur at the off-path unit that does not divide the architecture into two disconnected components.  Excluding these qubits breaks the Heavy-hex pattern discussed in Sec. \ref{sec:heavyhex}. However, we can still apply QFT pattern grid when missing off-path unit. All operations involving the missing unit could be ignored since the qubit does not even exist. For the off-path unit swapping with a missing unit, we can directly swap the target unit with the next available off-path unit. This long-distance off-path unit swapping would interact with all of the units along with the swapping path. To avoid violating the gate dependency, this long-distance swapping needs to wait for the on-path units to move to the right position first. 
This works for the non-faulty case where certain hanging nodes are omitted due to the turning points in the heavy-hex.

\textbf{\emph{Handling the error in Google Sycamore.}} To handle the faulty qubit in Sycamore, we could ignore two units. One contains that faulty qubit, called faulty unit. And another one adjacent to the faulty unit, called buffer unit. Unit swapping could be achieved with extra few costs for bypassing the faulty qubits. For the inter-unit interaction with one unit above and below two ignored units, we can move one to the buffer unit and proceed the inter-unit interaction.
\vspace{-12pt}
\section{Evaluation}
\label{sec:eval}
\paragraph{Metrics} We evaluate our QFT mapping approaches compared with other compilation approaches over different types of hardware architecture in several aspects. Specifically, we want to ask 3 major questions: (1) How long does it take to find the solution, (2) What is the quality of these solutions generated from various methods, and (3) Is the outcome generalizable for larger QFT sizes?

We write a simulator~\cite{anonymous_github} to verify the correctness of our outcome. We measure the quality of the compilation outcome, mainly based on circuit depth and gate count. Due to the noisy feature of quantum gate operations, smaller depth and fewer gate operations mean a lower possibility of being affected by external noise.

We present the compilation time in seconds.  If it cannot complete in 2 hours, we refer to it as ``time-out" in Table~\ref{table:expr}. 

Finally, we check the possibility of generalizing the compiled result into larger-size QFT circuits because it would be useful for us to exploit patterns in a small architecture (if possible) and reuse them in a larger but similar architecture. Our approach is definitely generalizable as described in previous sections. We will also check the generalizability of other compilers' outcomes. 

\textbf{Baseline and benchmark selection:} As for the baseline, we compare our approach against two state-of-the-art approaches: SATMAP~\cite{molavi+:micro22} and SABRE~\cite{li+:asplos19}.
SATMAP searches over the whole space and output optimal solutions (with respect to gate count) at the cost of long-running time.  SABRE does qubit mapping using a series of heuristics, making it possible to quickly output the potentially suboptimal results. 

\textbf{Architecture Backends:} We use three types of architectures: Google Sycamore, 2D grid and heavy-hex. Google Sycamore uses m by m configuration, where a total number of qubits is $N=m*m$. Since we reformed the architecture by merging two original rows into one unit, our unit size is $2m$.  For 2D grid, we use the square grid, where $N=m*m$. Each unit has m qubits. 

For heavy-hex, we unroll it to a line with dangling points \cite{weidenfeller+:arxiv22qaoa}. For every four qubits, it is a dangling qubit. There are $N*2/5$ units. One type of unit has a size of 4, and the other has a size of 1. We make $N$ a multiple of 5. It is possible that heavy-hex has a head and tail attached to the multiple of 5 qubits, making the total number of qubits not a multiple of 5. But these are the boundary cases that can be easily handled and would not asymptotically affect the overall complexity.

\subsection{Quality of Compilation Outcome}

\paragraph{Small-size QFT circuits}
 We test the output over several architecture types and for each type we would consider multiple configurations by setting the value m. 
We consider 3 major aspects. 
 First of all, how long does it take for each approach to outcome the results? The faster the better. 
 Besides, We consider the outcome quality as well, including circuit depth and gate count to complete the QFT. Completing the QFT with smaller depth and fewer SWAP gates is better.

Detailed results are in Table~\ref{table:expr}. Our approach has no compilation time since it is an analytical one. SATMAP outputs a good circuit depth and gate count, but it takes SATMAP quite a long time to find a solution due to the exponentially increasing search space. For instance, if we set the time-out threshold to be 2 hours, SATMAP cannot provide us the output when m is greater than 4 in the 2D grid and Sycamore. 

\begin{figure}[!t]
\begin{subfigure}[b]{0.23\textwidth}
    \centering
    \includegraphics[width=\textwidth]{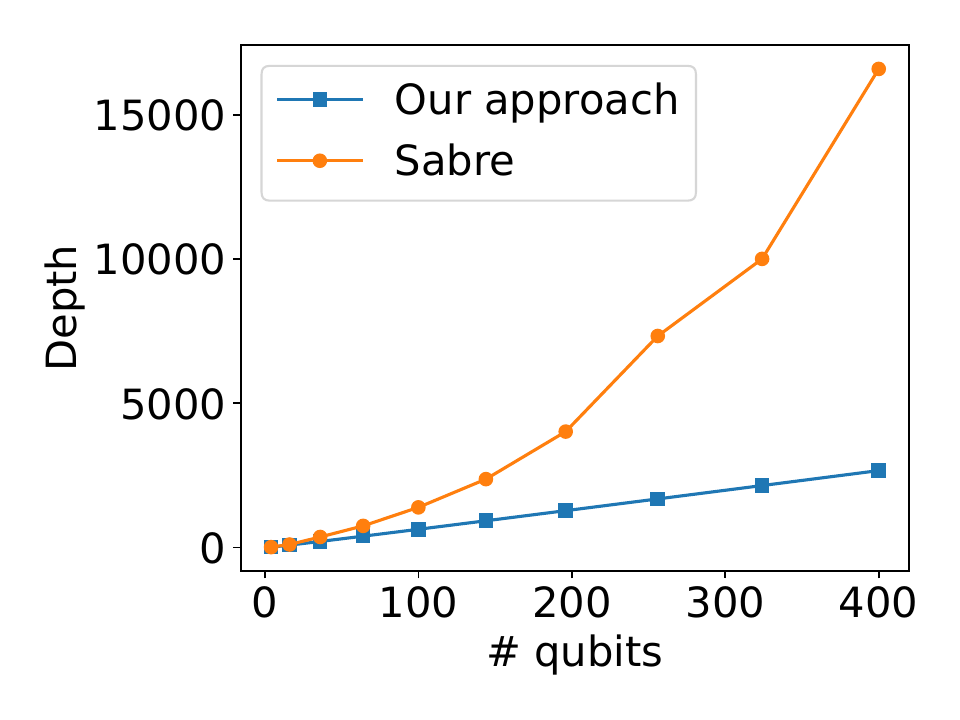}
    \vspace{-0.15in}
    \begin{small}
    \caption{Depth for Sycamore.}
    \end{small}
\end{subfigure}
\begin{subfigure}[b]{0.24\textwidth}
    \includegraphics[width=\textwidth]{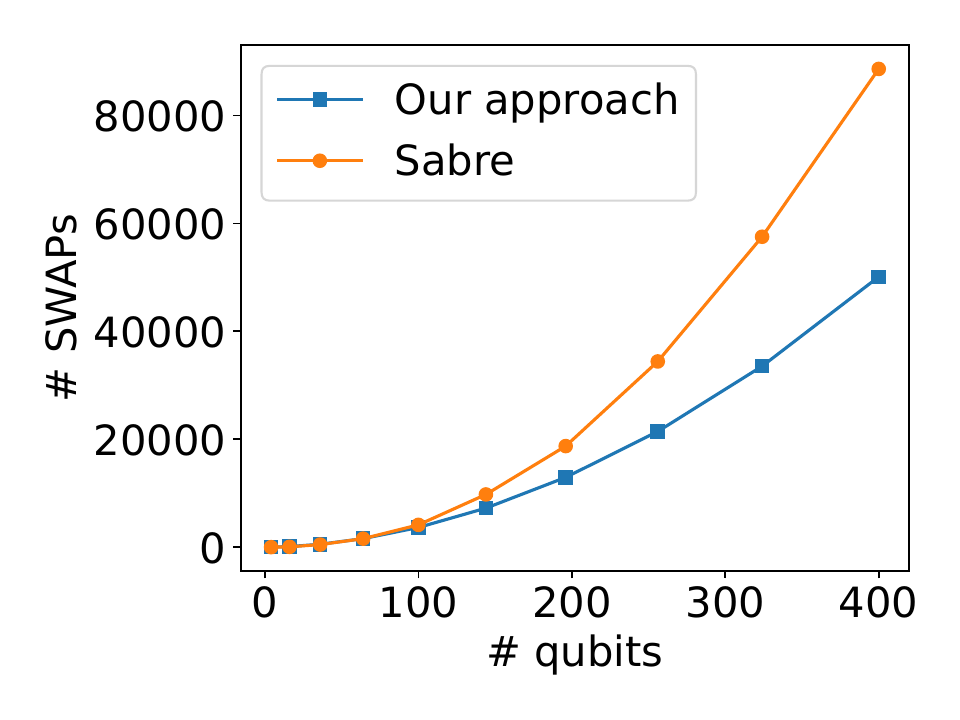}
    \vspace{-0.15in}
    \begin{small}
    \caption{\# SWAPs for Sycamore.}
    \end{small}
\end{subfigure}
\vspace{-0.25in}
\caption{Our approach vs. SABRE for Sycamore.\label{fig:Sycamore}}
\end{figure}

\begin{figure}[tb]
\begin{subfigure}[b]{0.23\textwidth}
    \centering
    \includegraphics[width=\textwidth]{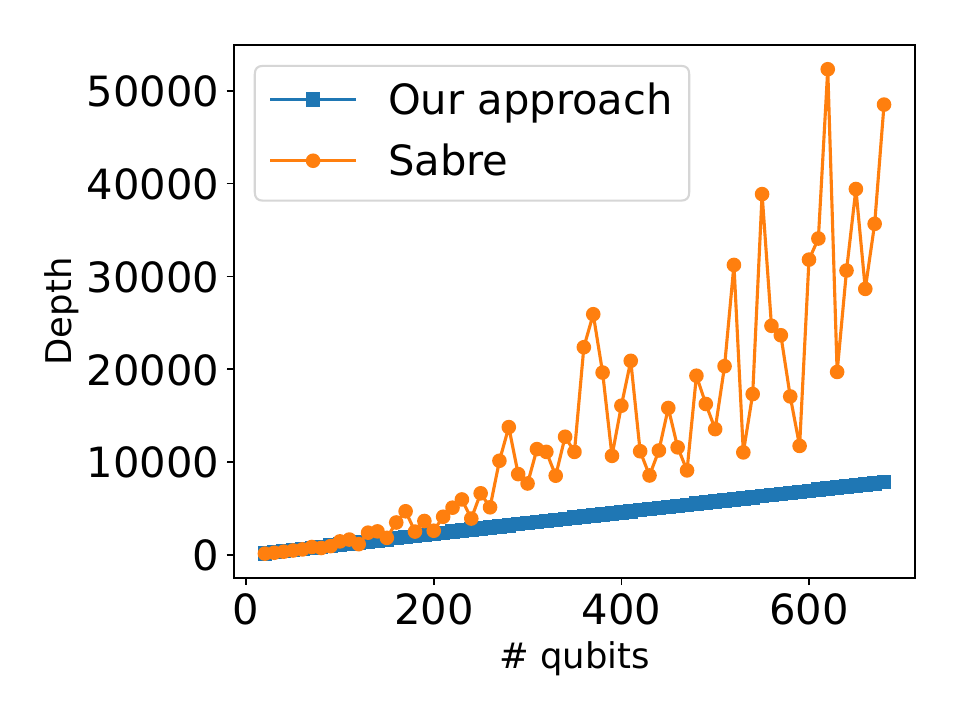}
    \vspace{-0.15in}
    \begin{small}
    \caption{Depth for Heavy-hex.}
    \end{small}
\end{subfigure}
\begin{subfigure}[b]{0.24\textwidth}
    \includegraphics[width=\textwidth]{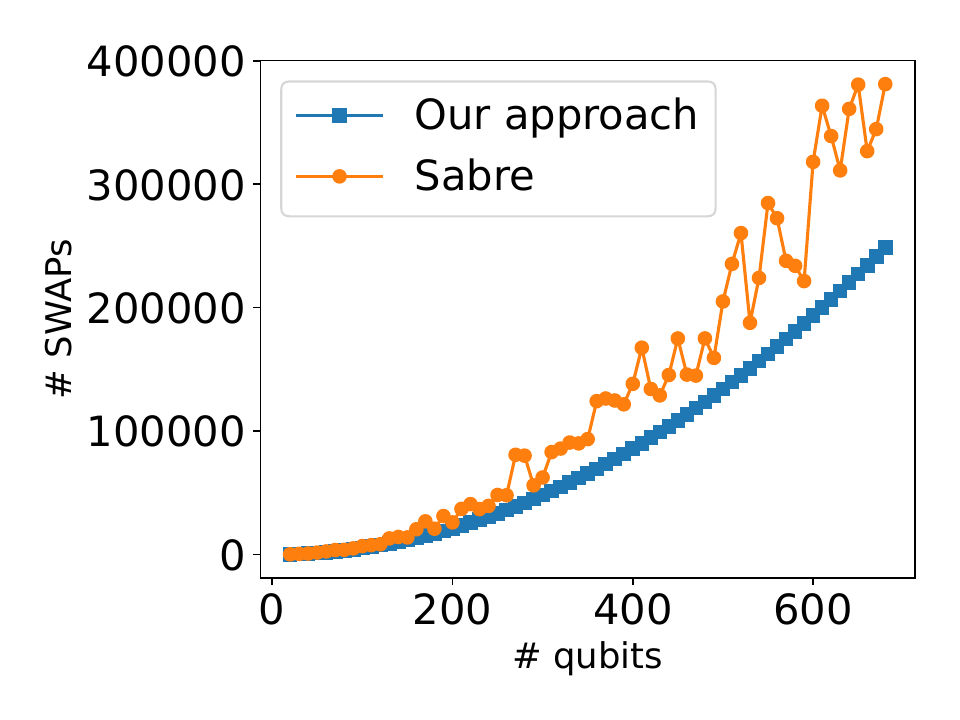}
    \vspace{-0.15in}
    \begin{small}
    \caption{\# SWAPs for Heavy-hex.}
    \end{small}
\end{subfigure}
\vspace{-0.25in}
\caption{Our approach vs. SABRE for Heavy-hex.\label{fig:heavy_hex}}
\vspace{-0.15in}
\end{figure}

\textbf{Large-size QFT circuits:}
SABRE can provide the outcome much faster than SATMAP at the cost of more usage of depth and SWAP gates. We compare our approach against SATMAP for 2D grid (Fig.~\ref{fig:NN}), Sycamore (Fig.~\ref{fig:Sycamore}), and heavy-hex (Fig.~\ref{fig:heavy_hex}) architecture.
As for depth, our approach can complete the QFT using much fewer cycles. As for \# SWAP, the general trend is that it is possible for SABRE to use fewer SWAP gates when the number of qubits is very small, but when the size increases, it will use more gates and depth. SABRE is not stable in certain cases, where it may have a smaller gate count for larger QFT sizes in the 2D grid and the heavy-hex architecture. 

\textit{Generally, we believe that our approach takes the best of worlds, being able to output reasonably good results in terms of depth and gate count quickly.}

\begin{figure}[htb]
\label{fig:NN}
\begin{subfigure}[b]{0.23\textwidth}
    \centering
    \includegraphics[width=\textwidth]{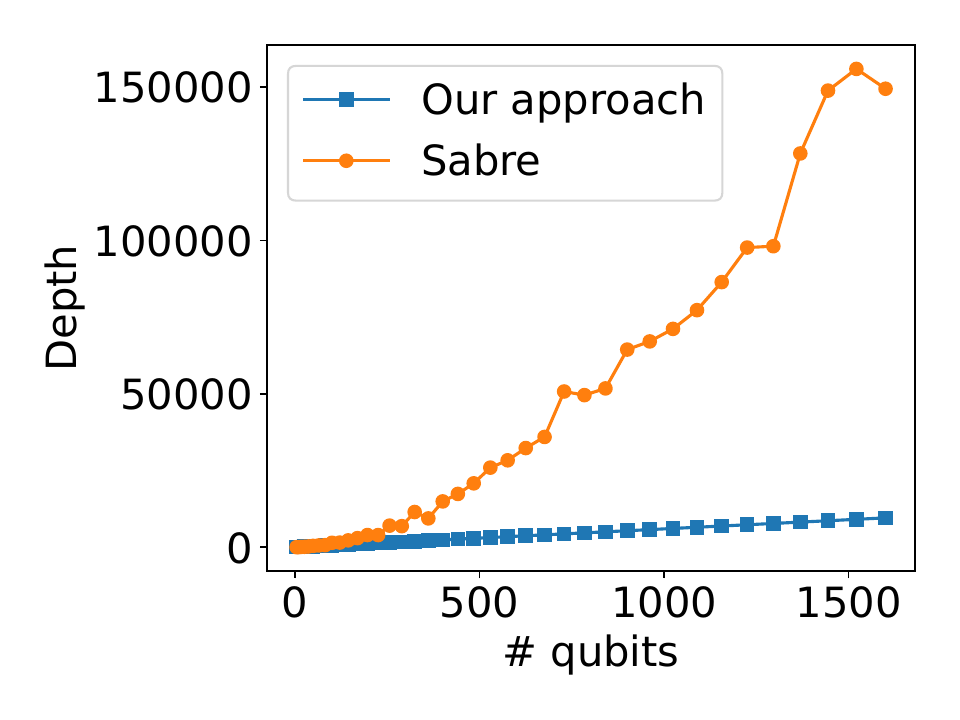}
    \vspace{-0.15in}
    \begin{small}
    \caption{Depth for 2D grid.}
    \end{small}
\end{subfigure}
\begin{subfigure}[b]{0.24\textwidth}
    \includegraphics[width=\textwidth]{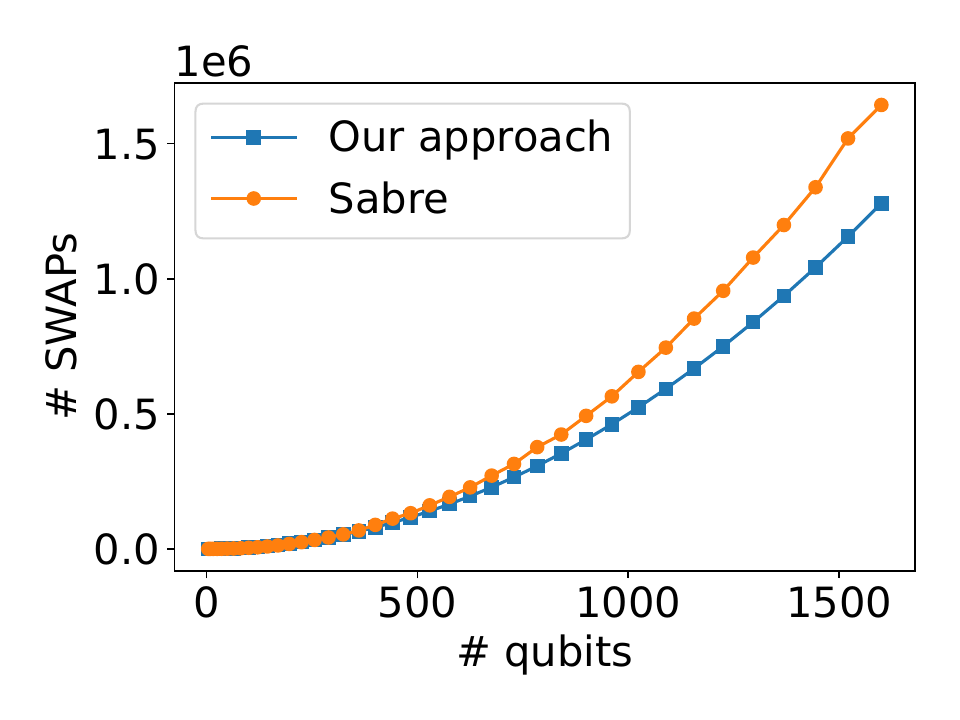}
    \vspace{-0.15in}
    \begin{small}
    \caption{\# SWAPs for 2D grid.}
    \end{small}
\end{subfigure}
\vspace{-0.25in}
\begin{small}
\caption{Our approach vs. SABRE for 2D grid.\label{fig:NN}}
\end{small}
\vspace{-0.1in}
\end{figure}

\begin{table}[]
\centering
\resizebox{0.48\textwidth}{!}{
\begin{tabular}{|l|l|ll|lll|lll|}
\toprule
\multirow{2}{*}{Architecture} & \multirow{2}{*}{\# qubits} & \multicolumn{2}{c|}{Our approach}    & \multicolumn{3}{c|}{SATMAP}                                        & \multicolumn{3}{c|}{SABRE}                                        \\ \cline{3-10} 
                      &                            & \multicolumn{1}{l|}{Depth} & \# SWAP & \multicolumn{1}{l|}{CT(s)}  & \multicolumn{1}{l|}{Depth} & \# SWAP & \multicolumn{1}{l|}{CT(s)} & \multicolumn{1}{l|}{Depth} & \# SWAP \\ \midrule
2D m*m                & 3*3                        & \multicolumn{1}{l|}{32}    & 33      & \multicolumn{1}{l|}{302.17} & \multicolumn{1}{l|}{35}    & 16      & \multicolumn{1}{l|}{0.28}  & \multicolumn{1}{l|}{32}    & 15      \\
2D m*m                & 4*4                        & \multicolumn{1}{l|}{66}    & 116     & \multicolumn{1}{l|}{TLE}     & \multicolumn{1}{l|}{N/A}      & N/A     & \multicolumn{1}{l|}{0.30}  & \multicolumn{1}{l|}{85}    & 71      \\
2D m*m                & 5*5                        & \multicolumn{1}{l|}{113}   & 295     & \multicolumn{1}{l|}{TLE}       & \multicolumn{1}{l|}{N/A}      & N/A     & \multicolumn{1}{l|}{0.34}  & \multicolumn{1}{l|}{140}   & 186     \\
2D m*m                & 6*6                        & \multicolumn{1}{l|}{172}   & 624     & \multicolumn{1}{l|}{TLE}       & \multicolumn{1}{l|}{N/A}      & N/A     & \multicolumn{1}{l|}{0.41}  & \multicolumn{1}{l|}{235}   & 403     \\
\midrule
m*m Sycamore          & 2*2                        & \multicolumn{1}{l|}{10}    & 6       & \multicolumn{1}{l|}{1.75}   & \multicolumn{1}{l|}{10}    & 3       & \multicolumn{1}{l|}{0.28}  & \multicolumn{1}{l|}{11}    & 3       \\
m*m Sycamore          & 4*4                        & \multicolumn{1}{l|}{81}    & 116     & \multicolumn{1}{l|}{TLE}       & \multicolumn{1}{l|}{N/A}      & TLE     & \multicolumn{1}{l|}{0.29}  & \multicolumn{1}{l|}{102}   & 62      \\
m*m Sycamore          & 6*6                        & \multicolumn{1}{l|}{208}   & 540     & \multicolumn{1}{l|}{TLE}       & \multicolumn{1}{l|}{N/A}      & N/A     & \multicolumn{1}{l|}{0.46}  & \multicolumn{1}{l|}{363}   & 484     \\
\midrule
Heavy-hex             & 2*5                        & \multicolumn{1}{l|}{42}    &  52    & \multicolumn{1}{l|}{439.79} & \multicolumn{1}{l|}{44}    & 37      & \multicolumn{1}{l|}{0.30}  & \multicolumn{1}{l|}{43}    & 36      \\
Heavy-hex             & 4*5                        & \multicolumn{1}{l|}{172}      &   189      & \multicolumn{1}{l|}{TLE}       & \multicolumn{1}{l|}{N/A}      & N/A     & \multicolumn{1}{l|}{0.31}  & \multicolumn{1}{l|}{134}    & 196      \\
Heavy-hex             & 6*5                        & \multicolumn{1}{l|}{288}      &     444    & \multicolumn{1}{l|}{TLE}       & \multicolumn{1}{l|}{N/A}      & N/A     & \multicolumn{1}{l|}{0.56}  & \multicolumn{1}{l|}{229}   & 523    \\
\bottomrule
\end{tabular}

}
\begin{small}
\caption{Our approach vs. satmap and sabre arcoss different architecture (CT: compilation time, TLE: timeout after 2h).
\label{table:expr}} 
\end{small}
\vspace{-20pt}
\end{table}

\subsection{Possibility of Generalizing It to Larger Size}
It is straightforward that our approach works for larger sizes. Our solution is written mostly in the form of affine loops in combination with modular functions. When the target architecture becomes larger, the code format is unchanged.

\begin{figure}[htb]
    \centering
    \includegraphics[width=0.46\textwidth]{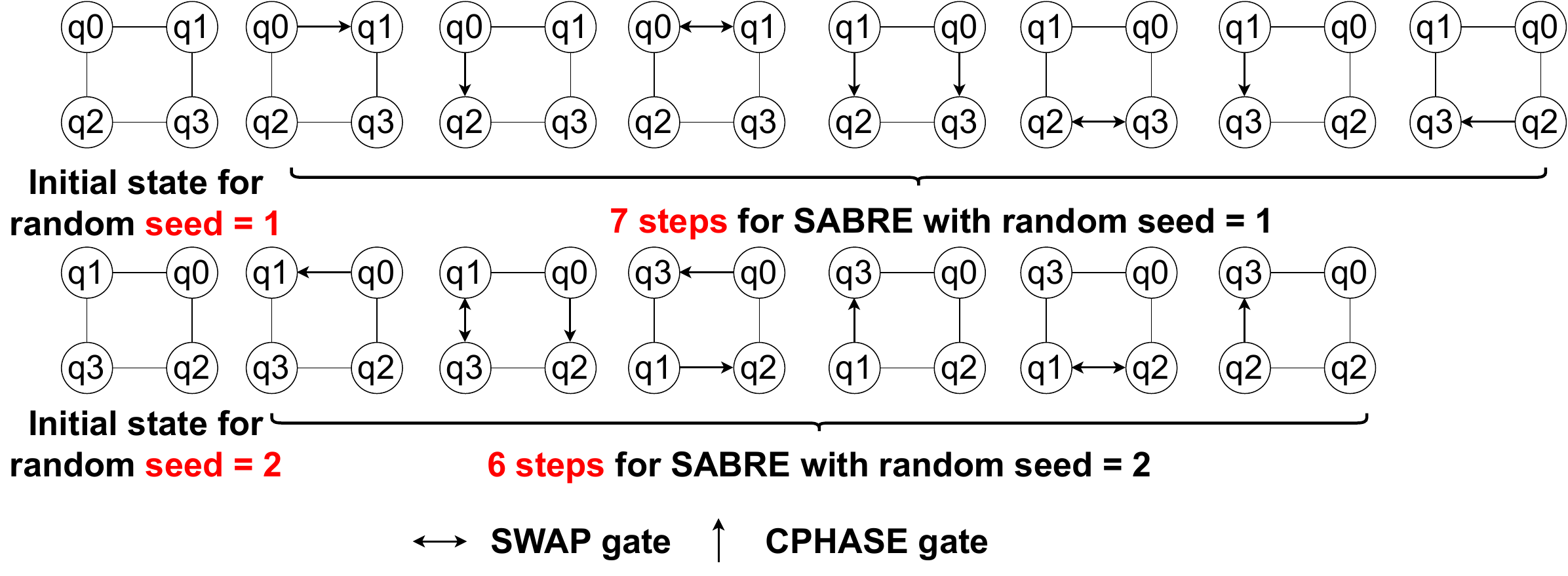}
    \vspace{-0.15in}
    \caption{Randomness in SABRE's output.}
    \label{fig:sabre}
    \vspace{-0.15in}
\end{figure}

Unfortunately, this is not the case for SABRE and SATMAP. SABRE tries to use lookahead to insert SWAPs that benefit not only the current layer but also for future layers. 
To make things more complicate, the outcome of SABRE is dependent on a random seed. Such randomness will often lead to dynamic outcome for each compilation.
For instance, Fig.~\ref{fig:sabre} presents the output from SABRE using different random seeds for QFT over 2 by 2 grid size. The outcome might vary from seeds to seeds in terms of initial mapping, gate operation order and even the final depth to complete QFT.
It is difficult for us to find any common patterns that can be reused for larger grid size.

Similarly, for the output generated by SATMAP, it is quite hard to generalize. The outcome for different paramater values could be completely different, making it impossible for us to learn something else from the output. Therefore, we could say that one output from SATMAP is useful specifically for one configuration. Since SATMAP mainly reduces gate count, it may not be able to find a structured solution. As pointed out earlier, we usually need to mirror the qubits after the mapping in order to ensure qubits are consecutively placed in a line before the next inter-unit operations, and this requires extra SWAPs, thus not gate optimal. Such a solution can only be encoded as structured loops, rather than a brute-force optimal solution for a particular size.


\section{Related Work}
\label{sec:relwork}

Many studies focus on qubit mapping for a general class of applications \cite{siraichi+:cgo18, li+:asplos19, zhang+:asplos21, molavi+:micro22, tan+:iccad20, tan+:iccad22}. A general-purpose compiler takes an arbitrary program and an arbitrary architecture as input, and produces a compiled circuit for this architecture. The issue with this approach is that every time the program size changes, 
for instance, the qubit number changes, 
the program needs to be recompiled. 
Our paper is different. We focus on domain-specific qubit mapping and do not require the compiler to recompile the program when the input size changes. We focus on QFT circuit mapping on 
important 
modern architectures. Our approach produces a linear-depth QFT circuit for Google and IBM superconducting architectures of arbitrary size. 

There are also other domain-specific compilers for various applications including quantum approximate optimization algorithms (QAOA)\cite{alam+:dac20, alam+:micro20, lao+:isca22},  variational quantum eigensolvers (VQE) \cite{li+:asplos22, li+:isca21}, and etc. 
For QFT, Maslov \etal   \cite{maslov+:physreva07} for the first time shows a linear time solution on the linear nearest neighbor (LNN) architecture. LNN represents a path of connected qubits. However, it is difficult to find a Hamiltonian path that connects all nodes in modern quantum architectures, limiting the applicability of this approach. Zhang \etal \cite{zhang+:asplos21} improved upon Maslov's \etal \cite{maslov+:physreva07} by discovering a linear-depth solution for a 2D grid with only two rows. However, architecture with only two rows as a 2 by l grid does not exist in modern architectures and is hard to generalize.

Leveraging program synthesis tools \cite{sketch} to do the compiler design for domain-specific applications has also existed. For instance, the packet-processing pipelines~\cite{rmt} designed via program synthesis method  (e.g., Domino~\cite{domino}, Chipmunk~\cite{chipmunk}, CaT~\cite{cat}). As far as we know, our work is the first that demonstrates the usefulness of program synthesis for the compiler design domain of quantum computing, for the QFT circuits.


\section{Conclusion}
\label{sec:conclusion}
We incorporate ``educated guess" into program synthesis solver to develop novel qubit mapping algorithms for QFT using affine loop format. 
Compared with state-of-the-art approaches, our qubit mapper can quickly output generalizable results with fewer usage of circuit depth and gate count.
We hope our results encourage 
engineers to adopt similar ideas to do qubit mapping for more emerging quantum hardware.

\bibliographystyle{plain}
\bibliography{allrefs}

\end{document}